%% file: preprint.tex
\documentstyle[emulateapj,apjfonts]{article}
\begin{document}

\begin{center}
{\bf The Hubble Space Telescope Key Project on 
the Extragalactic Distance Scale XXIV: \\
The Calibration of Tully--Fisher Relations and 
the Value of the Hubble Constant}

\lefthead{Sakai et al.}
\righthead{Tully--Fisher Relations and H0}

\end{center}

\def\Deg{\hbox{${}^\circ$\llap{.}}}
\def\Min{\hbox{${}^{\prime}$\llap{.}}}
\def\Sec{\hbox{${}^{\prime\prime}$\llap{.}}}

\def\kms{{\rm\thinspace km s$^{-1}$\thinspace \thinspace}}
\def\kmsmpc{{\rm\thinspace km s$^{-1}$ Mpc$^{-1}$\thinspace \thinspace }}
\def\iband{{\rm\thinspace I--band\thinspace \thinspace }}
\def\hband{{\rm\thinspace H--band\thinspace \thinspace}}
\def\hnot{{\rm\thinspace H$_0$\thinspace\thinspace }}

\begin{center}
Shoko Sakai$^{1}$,
Jeremy R. Mould$^{2}$,
Shaun M.G. Hughes$^{3}$,
John P. Huchra$^{4}$,
Lucas M. Macri$^{4}$,
Robert C. Kennicutt, Jr.,$^{5}$,
Brad K. Gibson$^{6}$,
Laura Ferrarese$^{7,8}$,
Wendy L. Freedman$^{9}$,
Mingsheng Han$^{10}$,
Holland C. Ford$^{11}$,
John A. Graham$^{12}$,
Garth D. Illingworth$^{13}$,
Daniel D. Kelson$^{11}$,
Barry F. Madore$^{14}$,
Kim Sebo$^{2}$,
N. A. Silbermann$^{14}$, and
Peter B. Stetson$^{15}$
\end{center}

\altaffiltext{1}{Kitt Peak National Observatory, National Optical Astronomy Observatories, Tucson AZ 85726, USA}

\altaffiltext{2}{Research School of Astronomy \& Astrophysics,
Institute of Advanced Studies, Australian National University, \\
Mount Stromlo Observatory Private Bag, Weston Creek, ACT 2611, Australia}

\altaffiltext{3}{Institute of Astronomy, Cambridge, UK}

\altaffiltext{4}{Harvard Smithsonian Center for Astrophysics, Cambridge, MA 02138, USA}

\altaffiltext{5}{Steward Observatories, University of Arizona, Tucson AZ 85721, USA}

\altaffiltext{6}{CASA, University of Colorado, Boulder, CO}

\altaffiltext{7}{California Institute of Technology, Pasadena CA 91125, USA}

\altaffiltext{8}{Hubble Fellow}

\altaffiltext{9}{Carnegie Observatories, Pasadena CA 91101, USA}

\altaffiltext{10}{Avanti Corp,  46871 Bayside Parkway, Fremont, CA 94538, USA}

\altaffiltext{11}{Johns Hopkins University and Space Telescope Science Institute, Baltimore, MD 21218}

\altaffiltext{12}{Department of Terrestrial Magnetism, Carnegie Institution of 
Washington, Washington, DC 20015, USA}

\altaffiltext{13}{Lick Observatory, University of California, Santa Cruz CA 95064, USA}

\altaffiltext{14}{IPAC, California Institute of Technology, Pasadena CA 91125, USA}

\altaffiltext{15}{Dominion Astrophysical Observatory, Victoria, BC V8X 4M6, Canada}

\medskip

\begin{abstract}
This paper presents the calibration of $BVRIH_{-0.5}$ Tully--Fisher relations
based on Cepheid distances to 21 galaxies within 25 Mpc, and 23 clusters within 
10,000 km s$^{-1}$.
These relations have been applied to 
several distant cluster surveys in order to
derive a value for the Hubble constant, H$_0$, mainly concentrating
on an I--band all--sky survey by Giovanelli 
and collaborators which consisted of total I magnitudes
and 50\% linewidth data for $\sim$550 galaxies in 16 clusters.
For comparison, 
we also derive the values of H$_0$ using surveys in B--band and V--band
by Bothun and collaborators, and in H--band by Aaronson and collaborators.
Careful comparisons with various other databases from literature suggest
that the H--band data, whose magnitudes are isophotal magnitudes 
extrapolated from aperture magnitudes rather
than total magnitudes, are subject to systematic uncertainties.
Taking a weighted average of the estimates of Hubble constants from
four surveys, we obtain H$_0$ = 71 $\pm$ 4 (random) $\pm$ 7 (systematic)
We have also investigated how various systematic uncertainties affect
the value of H$_0$ such as the internal extinction correction method used,
Tully--Fisher slopes and shapes, a possible metallicity dependence of
the Cepheid period--luminosity relation and cluster population 
incompleteness bias.

\end{abstract}

\section{Introduction}

The Tully--Fisher relation, a luminosity--linewidth correlation
for spiral galaxies, has become the most widely used method for
determining galaxy distances independently of redshifts.
Although the concept had been used variously, for example, by
\"Opik (1922) and Roberts (1969, 1975) over several decades, 
it was the groundbreaking paper by Tully \& Fisher (1977: hereafter TF77) 
which laid out its potential as a powerful extra-galactic distance indicator.
The Tully-Fisher (TF) relation is an empirical correlation between the 
absolute magnitude of a late-type spiral galaxy and its maximum 
rotational velocity.  
The latter quantity is usually  determined from
radio observations of the neutral hydrogen 21cm emission spectrum or from
the optical rotation curve and  is distance--independent.
The TF relation has been used not only to investigate the density and velocity
fields of the large--scale distributions of galaxies, but also to measure
the value of the Hubble constant.  Readers are referred to Dekel (1994),
Strauss and Willick (1995) and
Giovanelli (1997) for extensive reviews on these subjects.

After the first application of the TF relation using B--band photometry
of spiral galaxies (TF77), it was soon discovered that
there were mainly two advantages to using near--infrared
over visible--wavelength photometry: (1) the uncertainties due
to interstellar absorption in both our own Galaxy and in the observed
galaxy would be greatly reduced; and (2) the infrared emission is
dominated by late--type giants and thus there is no significant
dependence of mass--to--light ratio on galaxy morphology.
The first study to apply  the H--band (1.65\micron) -- HI linewidth TF relation was 
reported by Aaronson, Huchra \& Mould (1980), who used
a large--aperture H--band magnitudes to obtain distances to
the M81, Sculptor and M101 groups.
Aaronson and Mould (1983) later reported that in the infrared, 
the TF relation is morphology-independent, unlike in the blue
where there is a small dependence.
They also found that the slope in the relation is wavelength--dependent
with its value increasing from the blue to IR.
The scatter has been reported to be smaller for the infrared relation,
making it a better distance indicator.
Furthermore, the TF relation has been reported to be 
different for field and cluster
galaxies or for galaxies in different clusters (Sandage and Tammann 1984).

Major improvement in the TF relation has been achieved during the late
1980s using CCD images.  As a result, numerous all--sky
surveys of spiral galaxies were carried out which used the TF
relation to estimate their distances, to probe the density and
velocity fields of the large--scale structure of the Universe.
These include an r--band survey by Willick (1991) who used the 21cm
linewidths to measure the rotation velocity, and also by Courteau
(1992) who used the optical rotation curves.  Han (1991) prsented
an I--band TF all--sky survey, which concentrated on the galaxies
previously observed by Aaronson et al. (1982: hereafter AHM82) in 
their H--band survey.
Most recently, Giovaneli et al. (1997: hereafter G97) presented an
all--sky I--band survey of cluster galaxies out to $\sim$10,000
km s$^{-1}$.  Mathewson et al. (1995) also measured
I--band TF distances to clusters in the Southern hemisphere.
Although the data have not been published as of now, 
Willick (1999) has presented a peculiar--velocity study based
on his Las Campanas/Palomar Survey which is comprised of R--band
photometry and H$\alpha$ rotation curve observationas of spiral
galaxies out to 10,000 km s$^{-1}$.
Dale et al. (1999) also has presented the velocity field study
based on their I--band spiral--galaxy survey reaching as far out as
$\sim$25,000 km s $^{-1}$.
Courteau (1999) has also recently completed V and I--band photometric
and H$\alpha$ rotation--curve observations of spiral galaxies.

Although there is no firm physical basis for the TF relation,
several groups have tried to recover the TF physical parameters
using cosmological numerical simulations.
Some examples include that by Steinmetz and Navarro (1999)
who used high--resolution standard CDM simulations.
They reported that the slope of the TF relation can be
reproduced in a hierarchical clustering model naturally,
suggesting that the TF relation is a direct consequence
of the cosmological equivalence between the mass and circular
velocity, which was also the conclusion of Mo, Mao \& White (1998).
On the other hand, Eisenstein \& Loeb (1996) argued that 
because the scatter due to variations in formation histories
is larger than that observed, the TF relation is a consequence
of subsequent feedback processes which regulate the gas dynamics
and star formation in these galaxies.

The main objective of the Hubble Space Telescope Key Project on the
Extragalactic Distance Scale (H$_0$ Key Project) is to measure
the value of the Hubble constant to an accuracy of 10\%.  This is
achieved by first obtaining precise Cepheid distances to nearby spiral
galaxies and then by calibrating several secondary distance indicators.
The H$_0$ Key Project has now completed the observations of all of
its target galaxies and the Cepheid data and distances for each galaxy
have been presented as part of this series in papers I through XXIII.
This paper presents research and analysis which are part of our final
effort: the determination of the Hubble constant through the calibration
of a variety of secondary distance indicators.   In particular,
this paper focuses on the calibration of the optical and near--IR TF
relations using the Cepheid distances and the applications of the calibrated
relations to distant cluster samples to determine the value of the Hubble constant.
Other secondary distance indicators that are calibrated
as part of the H$_0$ Key Project are the surface brightness
fluctuation method (Ferrarese et al. 2000a), 
Type Ia supernovae (Gibson et al. 2000) and fundamental plane
(Kelson et al. 2000).
The measurements of the Hubble constant 
will then be combined together
in a consistent fashion in Mould et al. (2000).
The extrapolation from a local value of the Hubble constant
to the global value will be discussed in Freedman et al. (2000).

We have obtained new
BVRI surface photometry of most of the galaxies suitable for the absolute
calibration of the TF relation. These data are presented in Macri et al.(2000)
and summarized in Section 2.
The photometric data will be combined with the rotational velocity data
compiled from the literature to calibrate TF relations in these
four optical wavelengths and in H~band (Section~\ref{section:tfcalibration}).
We also address the values of the intrinsic dispersion
of the TF relations and the role played by a second parameter and
non--linearities.
The calibrated TF relations are then applied to
distant clusters,
mainly those from the I--band survey presented by Giovanelli et al. (1997a).
The I--band TF distances to those clusters and values of H$_0$ are 
presented in Section 4.
They are then compared with values obtained by examining surveys
at other wavelengths.
In Section 5, various systematics that could affect
the measurement of the H$_0$ are examined.  These include:
(1) consistency between the calibrators and cluster samples;
(2) dependence on the TF slopes and shapes;
(3) dependence on the internal extinction correction method used;
(4) metallicity dependence of the Cepheid period--luminosity relations;
(5) large--scale velocity field of the Universe;
and (6) cluster population incompleteness bias.
Finally in Section 6, our H$_0$ values are compared with previously derived
TF H$_0$ estimates.

\section{Data}
\label{section:data}

\subsection{Cepheid Distances}

In Table 1,
we list the spiral galaxies used in the TF analysis and the sources
of their Cepheid distances.
The list is comprised of : (1) 15 galaxies which were observed as part of the
H$_0$ Key Project, (2) 3 galaxies whose Cepheid distances were measured by other
groups using the HST, and (3) 3 galaxies with Cepheid distances determined from
ground--based observations.
Note that there are several galaxies, such as NGC~4321 (M100) and NGC~5457 (M101),
that have been observed by the Key Project but are not suitable for calibrating
the TF relation because their inclinations are too small.

The distances given in Table~1 were all taken directly from the references
as indicated.  However, the corresponding errors were re--estimated 
so that they are consistent from galaxy to galaxy,
as the understanding of uncertainties in Key Project Cepheid distances
have grown over the course of the project.
The details of which and how uncertainties were treated are
found in Ferrarese et al. (2000b).

\tablenum{1}

{\scriptsize
\def\h0kp{{\rm$H_0$ KP}}
\begin{deluxetable}{lcccc}
\tablecolumns{5}
\tablewidth{0pc}
\tablecaption{\bf Cepheid Distances}
\tablehead{
\colhead{Galaxy} &
\colhead{Type} &
\colhead{$m - M$ (mag)} &
\colhead{Reference} &
\colhead{Project}
}
\startdata
NGC 224   & 3  &  24.44 $\pm$ 0.10 &   Freedman 1990  & ground-based \cr
NGC 598   & 6  &  24.64 $\pm$ 0.09  &  Freedman 1990  & ground-based \cr
NGC 925   & 7  &  29.84 $\pm$ 0.08  &  Silbermann et al. 1996  & H$_0$ KP \cr
NGC 1365  & 3  &  31.39 $\pm$ 0.20  &  Silbermann et al. 1999 / Ferrarese et al. 2000 & H$_0$ KP \cr
NGC 1425  & 3  &  31.81 $\pm$ 0.06  &  Mould et al. 1999 / Ferrarese et al. 2000 & H$_0$ KP \cr
NGC 2090  & 5  &  30.45 $\pm$ 0.08  &  Phelps et al. 1998 & H$_0$ KP \cr
NGC 2403  & 6  &  27.51 $\pm$ 0.24  &  Madore \& Freedman 1991 & ground--based \cr
NGC 2541  & 6  &  30.47 $\pm$ 0.08  &  Ferrarese et al. 1998 & H$_0$ KP \cr
NGC 3031  & 2  &  27.80 $\pm$ 0.08  &  Freedman et al. 1994 & H$_0$ KP \cr
NGC 3198  & 5  &  30.80 $\pm$ 0.06  &  Kelson et al. 1999 & H$_0$ KP \cr
NGC 3319  & 3  &  30.78 $\pm$ 0.12  &  Sakai et al. 1999 & H$_0$ KP \cr
NGC 3351  & 3  &  30.01 $\pm$ 0.08  &  Graham et al. 1997 & H$_0$ KP \cr
NGC 3368  & 2  &  30.20 $\pm$ 0.10  &  Tanvir et al. 1995/Gibson et al. 2000 & HST \cr
NGC 3621  & 7  &  29.13 $\pm$ 0.11  &  Rawson et al. 1997 & H$_0$ KP \cr
NGC 3627  & 3  &  30.06 $\pm$ 0.17  &  Saha et al. 1999/Gibson et al. 2000 & HST \cr
NGC 4414  & 5  &  31.41 $\pm$ 0.10  &  Turner et al. 1998 & H$_0$ KP \cr
NGC 4535  & 5  &  31.10 $\pm$ 0.07  &  Macri et al. 1999 / Ferrarese et al. 2000 & H$_0$ KP \cr
NGC 4536  & 4  &  30.95 $\pm$ 0.08  &  Saha et al. 1996/Gibson et al. 2000 & HST \cr
NGC 4548  & 3  &  31.04 $\pm$ 0.23  &  Graham et al. 1999 & H$_0$ KP \cr
NGC 4725  & 2  &  30.57 $\pm$ 0.08  &  Gibson et al. 2000 / Ferrarese et al. 2000 & H$_0$ KP \cr
NGC 7331  & 3  &  30.89 $\pm$ 0.10  &  Hughes et al. 1998 & H$_0$ KP \cr
\enddata
\label{table:cephdist}
\end{deluxetable}
}

\subsection{Photometry of Tully--Fisher Calibrators}

We summarize briefly in this section the surface photometry observations
and results, as this subject is presented in full detail by 
Macri et al. (2000).
$BVRI$ observations of the Tully-Fisher calibrator galaxies were conducted
between March 1994 and May 1999 using the Fred L. Whipple Observatory 1.2-m
telescope and the Mount Stromlo and Siding Springs Observatories 1-m
telescope.  %Most of the objects were observed from FLWO, while MSSSO was used
Exponential
disks were fitted to the surface brightness profiles, following
Han (1992), to extrapolate total magnitudes. Both the profile fitting and
the determination of the inclination were done interactively, in a manner
similar to Giovanelli et al. (1997a).

For the IR data, we use $H_{-0.5}$ magnitudes presented
by Aaronson et al. (1982), since with the exception of some work
by Bernstein et al. (1994), that is all that is currently available 
in the literature, in terms of a consistent set of magnitudes between 
the calibrators and cluster galaxies.

\subsubsection{Corrections Applied to the Photometric Data}

For the TF application,
the photometric data were then corrected for Galactic extinction ($A_{G,\lambda}$),
internal extinction ($A_{int,\lambda}$) and k--correction.
The Galactic extinction values were estimated using the 
100$\mu$m  maps reprocessed from IRAS/ISSA and COBE/DIRBE
data by Schlegel et al. (1998).
The $I-$band cosmological k--term was adopted from Han (1992): $k_I = 0.16 z$.

The amount of internal extinction in spiral galaxies and its
correction methods still remains a very controversial issue.
Several papers on this issue include Holmberg (1958), Valentijn (1990),
Han (1992), Giovanelli et al. (1994), 
Burstein, Willick \& Courteau (1994), Tully et al. (1998). 
For the internal extinction corrections, we have decided to apply those
derived by Tully et al. (1998: hereafter T98), which are functions of both 
major--to--minor axes ratio $(a/b)$ and linewidth, and are expressed as:
\begin{equation}
A^{T98}_{int,BRIK} = \gamma_{\lambda} \log (a/b),
\end{equation}
where $\gamma_{\lambda}$ is a function of linewidth:
\begin{equation}
\gamma  = \left\{ \begin{array}{ll}
	(1.57 + 2.75 (\log (W_{20\%}) - 2.5)) &  \mbox{B} \\
	(1.15 + 1.88 (\log (W_{20\%}) - 2.5)) &  \mbox{R} \\
	(0.92 + 1.63 (\log (W_{20\%}) - 2.5)) &  \mbox{I} \\
	(0.22 + 0.40 (\log (W_{20\%}) - 2.5)) &  \mbox{K} \\
	\end{array}
\right.
\end{equation}
The idea behind the dependence on linewidth is
that the light needs to travel further on average in a larger galaxy,
thus requiring a larger extinction correction.
However, it is noted that careful examination by Willick et al. (1996)
revealed no dependence of the internal extinction on linewidth or lumiosity.

In Appendix A, we review two other expressions for
internal extinction derived
by Han (1992) and Giovanelli et al. (1994).  The main difference in
these two studies is that the extinction corrections are dependent
on the morphological types of galaxies.
Mainly because the uncertainty in the morphological classification of distant
galaxies is large, we have decided to use the T98 formulae in
this paper.
Later in Section 5.7, we will examine whether the value of H$_0$ 
depends on the internal correction method.  

The T98 corrections in Equation 1 include no independent formulae
for the V and H bands.
For these, we adopt the ratio $A_V = 1.5A_I$ and $A_H = 0.5A_I$, which
are reasonable values obtained by interpolating the T98 ``extinction
law'' (Equations 1 and 2).
Note that these ratios are slightly smaller than those derived by
Cardelli, Clayton and Mathis (1989).
This is to be expected since Cardelli et al.  used a fixed value for the
optical path length, when in reality, blue light is attenuated more quickly
than red, and thus different optical path lengths are sampled at
different wavelengths (shorter in the blue than in the red) (Han 1992, T98).

Putting all the corrections together, the corrected 
apparent magnitude is then expressed as:
\begin{equation}
m^c_{\lambda} = m_{\rm obs,\lambda} - A_{G,\lambda} - A_{int,\lambda} 
- k_{\lambda} 
\end{equation}

Note that the essential point for present purposes is to adopt {\it consistent}
corrections between the calibrator and the distance galaxy samples. 
We emphasize that the corrections 
do not need to be optimum; they need to be consistent.

\subsubsection{Inclination}

Accurate measurement of inclination angles is %extremely 
crucial in the application of the TF relation since it is possibly
the largest source of error in both magnitudes and linewidths.
We measured the inclination in all four bands (BVRI),
which agreed with each other well within 1$\sigma$
errors (see Macri et al. 2000 for details),
which will be used in the rest of the paper to calibrate
the TF relations and also to measure the value of H$_0$.

In Appendix B, we summarize the inclination information compiled
from previously published data for each galaxy in detail.
In some cases, such as NGC 1365 or NGC 4725, the kinematical inclination 
angle is quite different from
the photometric inclination angles that we obtained.
Although this could potentially lead to 
uncertainties in the corrected magnitudes and linewidths, 
this should not introduce a bias in our analysis:
for the distant cluster surveys, which will be used
to measure the value of the H$_0$, the inclinations of galaxies were
measured strictly from the imaging, either on
photographic plates or on CCDs.  Because our main objective is
to create a calibration sample that is consistent with the
cluster surveys, only the photometric inclination
angles (from Macri et al. 2000) will be considered for further analysis.

\subsection{Linewidths}

Although linewidths have been measured and published for all the Key Project
galaxies selected as calibrators for the Tully-Fisher relation, we have 
endeavored wherever possible to re-measure the various linewidth definitions
directly from the profiles, in order to check for any anomalies.
We are  grateful to Martha Haynes for allowing us access to the profiles of 
the  KP galaxies which were observed as part of the Giovanelli et al 
(1997a) database.
In addition, several 21cm line profiles were digitally scanned
from the literature.
We review here only briefly the linewidth database, as the details
are explained in Macri et al. (2000).

In order to avoid systematic effects when applying our
Cepheid calibration to various distant data sets, it is important that we 
use the applicable linewidth definition. 
The 20\% linewidths, $W_{20}$, are often used in the TF applications, which
are the width in km s$^{-1}$ of the HI profile measured at 
20\% of the peak of the  HI flux, where for a 2--horned profile this is taken 
as the mean of each horn's peak flux. 
Another commonly used linewidth measurement is the 50\%
width, which is, for example, used in the I--band data set of
G97.
The $W_{50}$ is the width at 50\% of each of the horn maxima 
(or if the profile is Gaussian-shaped, then the width at 50\% of 
the single peak). 
The potential advantage of measuring the linewidth at the 50\% flux level is 
that it is less likely to be affected by noise, especially for low S/N 
profiles.  However, the G97 $W_{50}$'s are
measured from an  interpolation between the levels at 15\% and 85\% of the 
peak flux.  For  unusual profiles, this can result in underestimated profile 
widths.

As all the KP galaxies are relatively nearby, most
have very high S/N profiles.  Many of them are also very large and 
overfill the telescope's beamsize.  In these cases, to avoid
systematically  underestimating the profile widths,  the observers have claimed to
have been careful to ensure that a sufficient number of pointings were 
obtained.

Most galaxies have the classic steep-sided twin-horned profiles of normal 
edge-on spirals. 
Unfortunately, there are exceptions.  
Galaxies with different profiles are NGC~3031 (which is 
distorted by a tidal interaction with M82=NGC~3034), NGC~3319 (which has a 
low-level pedestal), and NGC~3368 (which has a non-symmetric two-horned profile).
The $W_{50}^G$ widths are particularly
sensitive to both low S/N profiles and pedestal features (which affect
the positions of the 15\% level) and unusual profiles (as the 85\% level can 
be far from the end of the steep edge), the latter leading to systematically 
underestimated widths.  These cases are flagged by larger uncertainties.
%we have not attempted to correct for any systematic effects.  

\subsubsection{Corrections Applied to the Linewidths}

We apply the following corrections to the raw linewidth data.
First,  linewidths were corrected for the inclination angle ($i$)
and also for the redshift ($z$) effect following:
\begin{equation}
W^c = W (\sin i)^{-1} (1+z)^{-1}.
\end{equation}
The inclination angle is derived from the eccentricity measured
on the CCD frames via:
\begin{equation}
i = \cos^{-1}\sqrt{\frac{(b/a)^2 - q_0^2}{1 - q_0^2}},
\end{equation}
where $q_0$ is the intrinsic minor--to--major axis ratio
of edge--on spiral galaxies for which we adopt
$q_0=0.13$ for $T > 3$ and $q_0=0.20$ for other types.
No turbulence corrections are applied in this paper as the physical
basis for this effect still remains ambiguous, except in estimating
the internal extinction corrections (in Section 3.2).
Originally, one of the reasons for applying the turbulence correction was to straighten
the TF relation, in addition to some theoretical preconceptions.
Furthermore, as we will discuss in later sections, the calibration 
galaxies for our multi--wavelength TF relations mostly populate
the region where the linewidths exceed $\sim$300 kms$^{-1}$.
Since the turbulence correction is very small and constant for luminous galaxies
($\log W \geq 2.5$), applying
a turbulence correction would have no significant effect on our determination of
H$_0$.

\subsection{Data Presentation}   

In Table 2,
we list all of the photometric, inclination and kinematical data used
in calibrating the TF relation.
The columns for Table~2 are: 
(1) name of the galaxy;
(2)--(6) $BVRI$ total and $H_{-0.5}$ aperture magnitudes corrected for Galactic extinction
and internal extinction, as described in Section 2.2.  
For raw, uncorrected magnitudes, see Macri
et al. (2000).  The photometric data for NGC~224, NGC~598, NGC~2403 and NGC~3031
were taken directly from Pierce and Tully (1992);
(7) logarithmic 20\% linewidth corrected for inclination (using the mean inclination) 
and redshift;
(8) logarithmic 50\% linewidth corrected for inclination and redshift;
(9) photometric I--band inclination; and
(10) the mean photometric inclination which is used to calculate the internal
extinction, and also to correct for inclination for the linewidths.
We list here the I--band photometric inclination angles,
since the strategy is to bring our photometry/linewidth
data to the same system as the distant cluster surveys.
As will be described in detail later on, the I--band survey by G97
estimated the galaxy inclinations from their I--band images.
We note, however, that in most cases, the $BVRI$ photometric
inclinations agree with each other extremely well, within 1$\sigma$ errors.

{\scriptsize
\tablenum{2}
\begin{deluxetable}{lccccccccc}
\tablecolumns{12}
\tablewidth{0pc}
\tablecaption{\bf Photometric and Kinematical Data for Tully--Fisher Calibrators} 
\tablehead{
\colhead{Galaxy} &
\colhead{$B^c_T$\tablenotemark{a}} &
\colhead{$V^c_T$\tablenotemark{a}} &
\colhead{$R^c_T$\tablenotemark{a}} &
\colhead{$I^c_T$\tablenotemark{a}} & 
\colhead{$H_{-0.5}^c$\tablenotemark{a}} &
\colhead{$\log W_{20}^c$\tablenotemark{b}} &
\colhead{$\log W_{50}^c$\tablenotemark{b}} &
\colhead{$i_I$\tablenotemark{c}} &
\colhead{$\bar{i}$\tablenotemark{c}}  \\
\colhead{} &
\colhead{mag} &
\colhead{mag} &
\colhead{mag} &
\colhead{mag} &
\colhead{mag} &
\colhead{kms$^{-1}$} &
\colhead{kms$^{-1}$} &
\colhead{deg} &
\colhead{deg} 
}
\startdata
N224\tablenotemark{d} &   -21.37 (19)  &    \nodata   &   -22.47 (18)  &   -23.04 (18)  &  -23.83 (11)   & 2.744(28) &  2.703(30) & 78(2) & 78(2) \cr 
N598\tablenotemark{d} &   -18.57 (18)  &    \nodata   &   -19.39 (18)  &   -19.80 (18)  &  -20.31 (11)   & 2.397(74) &  2.357(82) & 54(2) & 54(2) \cr
N925                  &   -19.59 (29)  &  -19.95 (16) &   -20.24 (16)  &   -20.59 (28)  &  -21.16 (10)   & 2.420(49) &  2.384(52) & 56(1) & 56(2) \cr
N1365                 &   -21.86 (37)  &  -22.30 (12) &   -22.71 (17)  &   -23.38 (12)  &  -24.30 (11)   & 2.682(35) &  2.626(32) & 62(3) & 55(2) \cr
N1425                 &   -20.96 (17)  &  -21.42 (08) &   -21.82 (07)  &   -22.38 (08)  &  -23.07 (08)   & 2.621(41) &  2.599(43) & 64(1) & 63(1) \cr
N2090                 &   -19.83 (11)  &  -20.20 (10) &   -20.67 (10)  &   -21.26 (11)  &  -21.95 (10)   & 2.501(35) &  2.493(38) & 64(1) & 67(2) \cr
N2403\tablenotemark{d}&   -19.11 (29)  &    \nodata   &   -19.93 (28)  &   -20.32 (28)  &  -21.14 (25)   & 2.480(59) &  2.457(62) & 58(2) & 58(2) \cr
N2541                 &   -18.72 (18)  &  -19.09 (13) &   -19.37 (10)  &   -19.82 (13)  &  -20.24 (10)   & 2.370(49) &  2.361(52) & 58(2) & 62(1) \cr
N3031\tablenotemark{d}&   -20.64 (17)  &    \nodata   &   -21.87 (17)  &   -22.43 (17)  &  -23.54 (10)   & 2.719(34) &  2.654(39) & 58(2) & 58(2) \cr
N3198                 &   -20.28 (08)  &  -20.75 (08) &   -21.07 (07)  &   -21.55 (08)  &  -22.22 (08)   & 2.531(32) &  2.506(34) & 71(2) & 68(2) \cr
N3319                 &   -19.34 (14)  &  -19.69 (14) &   -19.94 (13)  &   -20.32 (14)  &  -20.44 (13)   & 2.405(48) &  2.364(51) & 60(1) & 58(1) \cr
N3351                 &   -19.78 (11)  &  -20.56 (09) &   -21.00 (09)  &   -21.61 (09)  &  -22.56 (10)   & 2.586(47) &  2.577(42) & 45(2) & 45(2) \cr
N3368                 &   -20.47 (14)  &  -21.26 (12) &   -21.73 (11)  &   -22.29 (11)  &  -23.41 (11)   & 2.674(36) &  2.631(34) & 51(1) & 49(1) \cr
N3621                 &   -19.48 (12)  &  -20.00 (12) &   -20.50 (12)  &   -20.99 (12)  &  -21.84 (12)   & 2.499(35) &  2.471(38) & 64(1) & 66(1) \cr
N3627                 &   -21.10 (18)  &  -21.73 (18) &   -22.12 (18)  &   -22.63 (18)  &  -23.50 (18)   & 2.626(26) &  2.694(35) & 65(1) & 65(1) \cr
N4414                 &   -20.88 (13)  &  -21.57 (11) &   -21.98 (11)  &   -22.59 (11)  &  -23.65 (11)   & 2.743(39) &  2.562(39) & 46(2) & 46(2) \cr
N4535                 &   -20.80 (10)  &  -21.38 (09) &   -21.61 (08)  &   -22.27 (08)  &  -22.72 (09)   & 2.586(38) &  2.538(32) & 49(1) & 51(1) \cr
N4536                 &   -20.49 (12)  &  -21.05 (10) &   -21.39 (09)  &   -21.95 (13)  &  -22.79 (09)   & 2.562(30) &  2.695(31) & 68(1) & 69(1) \cr
N4548                 &   -20.36 (24)  &  -21.04 (23) &   -21.47 (23)  &   -22.13 (23)  &    \nodata     & 2.617(46) &  2.729(25) & 38(1) & 39(1) \cr
N4725                 &   -21.27 (10)  &  -21.85 (09) &   -22.20 (09)  &   -22.74 (09)  &  -23.65 (10)   & 2.671(26) &  2.564(30) & 51(1) & 62(1) \cr
N7331                 &   -21.58 (12)  &  -22.30 (11) &   -22.73 (11)  &   -23.29 (11)  &  -24.66 (11)   & 2.746(21) &  2.576(45) & 68(2) & 69(2)
\enddata								    
\tablenotetext{a}{The format of magnitudes is such that $-21.65(13)$ is equivalent to $-21.65 \pm 0.13$.}
\tablenotetext{b}{The format of $\log W$ is such that $2.682(35)$ is equivalent to $2.682 \pm 0.035$.}
\tablenotetext{c}{The inclination errors are shown in brackets.}
\tablenotetext{d}{Magnitudes from Pierce and Tully 1992}
\label{table:tfdata}
\end{deluxetable}					
}

\smallskip
\smallskip

\section{Calibration of Multi--Wavelength Tully--Fisher Relations}
\label{section:tfcalibration}

In this section, we derive TF calibrations using two independent
methods.  The first is to obtain a consistent set of $BVRIH_{-0.5}$
TF relations such that we can compare and examine the dispersions and 
the role of second
parameters.  This is done by using only the calibrator galaxies
with directly--measured Cepheid distances.  The 20\% linewidths and the
mean photometric inclination angles estimated from $BVRI$ measurements
will be used in this case.
The second method is to derive TF calibrations that
can actually be applied to the distant cluster surveys to
estimate H$_0$.  
This is done by using appropriate linewidth and inclination
angles, and also incorporating the cluster galaxy data to calculate the TF slopes.

\subsection{TF Calibration using Nearby Calibrator Galaxies Only}

First, we present $BVRIH_{-0.5}$ TF relations for 20\% linewidth,
using the mean of the $BVRI$ photometric inclination angles
(Col 10 of Table 2), derived from 21 nearby galaxies with Cepheid
distances only.
While the optical photometric data, $BVRI$, are all total magnitudes,
the H--band magnitude,
H$_{-0.5}$ is an aperture magnitude, extrapolated to the radius equivalent
to half the total blue light of the galaxies.
Slopes and zero points are determined using bivariate linear fits, 
minimizing errors in both $\log W^c$ and $M^c$: 

\begin{equation}
B^c_T = -7.85 (\pm 0.71) (\log W^c_{20} - 2.5) - 19.70 (\pm 0.11)
\hskip 0.25cm 
\end{equation}
\begin{equation}
V^c_T = -8.81 (\pm 0.82) (\log W^c_{20} - 2.5) - 20.27 (\pm 0.12)
\hskip 0.25cm 
\end{equation}
\begin{equation}
R^c_T = -8.68 (\pm 0.71) (\log W^c_{20} - 2.5) - 20.60 (\pm 0.11)
\hskip 0.25cm 
\end{equation}
\begin{equation}
I^c_T = -9.21 (\pm 0.75) (\log W^c_{20} - 2.5) - 21.09 (\pm 0.12)
\hskip 0.25cm 
\end{equation}
\begin{equation}
H^c_{-0.5} = -11.03 (\pm 0.86) (\log W^c_{20} - 2.5) - 21.74 (\pm 0.14)
\hskip 0.25cm 
\end{equation}

In Figure 1, the above TF relations are plotted.
The {\it observed} dispersions of above relations are $\sigma = 0.45, 0.37, 0.35,
0.37$ and $0.36$ for BVRI and H$_{-0.5}$ respectively,
which are combinations of observational errors and intrinsic scatter
of the Tully--Fisher relations themselves.  
The latter will be discussed in Section 3.4.
Errors in $\log W$ and absolute magnitudes were determined as follows.
The linewidth error is a combination of two sources:
the observational error in the linewidth measurement and the uncertainty
in the inclination angle.
For absolute magnitudes, their uncertainties come from a few sources including
the uncertainty in the Cepheid distance modulus (random only, typically $\sim$0.10 mag), 
observational photometric error (typically $\sim$0.10 mag),
the error in internal extinction law which is associated with the error in the
inclination angle, and the uncertainty in the Galactic extinction ($\sim$0.02 
mag in I--band).

We have also calculated the slope and zero point of each TF relation using
the direct and inverse fits for each wavelength as a consistency check,
in which the fits were determined by minimizing the errors in magnitudes
and linewidth only respectively.
Schechter (1980) suggested that the average value of $\log W$ at a constant
magnitude does not depend on whether the sample is distance-- or magnitude--limited.  
That is, the inverse TF relation is insensitive to the incompleteness
bias.  Based on this, some authors have preferred to use the inverse
TF relation to derive cluster distances (e.g. Tully 1999).
Although the slopes do change slightly as expected depending
on what fit is used, the zero point remains well within 1$\sigma$.
As will be shown in Section 5.2.1, the TF slope uncertainty has a very
little effect on the H$_0$ value.
In this paper, we will use the bivariate fits only.
Several articles have investigated the possibility of a TF relation being
a quadratic relation (cf: Mould, Han and Bothun 1989).  We do not use
quadratic TF relations in the main analyses of this paper, however, interested
readers are referred to Appendix C for further discussion.

\subsection{TF Calibration using Both the Calibrators and Cluster Galaxies}

We can also calibrate TF relations by combining the local calibrator
galaxies and cluster survey galaxies.
These independent samples should supplement each others' 
advantages and disadvantages.
The set of equations (6--10) will provide a consistent set of %sample of
TF calibrations which can be used to investigate several
fundamental questions pertaining to the TF correlation itself,
which will be discussed in the following sections.
However, in order to fulfill the original goal of this project,
to measure the value of H$_0$, we must in addition derive TF equations 
that are exactly on the same photometric--linewidth system as the distant cluster surveys.
Thus, a second set of the $I$ TF relations was derived using slightly different methods.

The G97 I--band survey used I--band photometric inclinations.
In determining the zero points and slopes, we take advantage
of the all--sky survey data that are available.  Calibrating the TF relations
using only local galaxies with Cepheid distances poses two problems.
This sample, which consists of only 21 galaxies, is not statistically
large enough to constrain the slope very well, and errors in the
individual galaxy distances contribute significantly to the final 
dispersion.  To minimize the first problem,
we iteratively determine the slope and zero point using both the distant
cluster sample and the local calibrators as follows: 
(1) the slope and zero point are calculated using the calibrator 
sample only.  (2) Then distances
to distant cluster galaxies (e.g. the G97 \iband sample) are determined 
using the calibration from (1), such that all the clusters are put on the same 
$(\log W - 2.5) - M_I$ plane.  Here, the cluster sample galaxies are corrected
for the cluster incompleteness bias (which will be discussed in detail
in Section 5.4).
(3) The slope is determined using this combined cluster sample.
(4) The zero point is then estimated by minimizing the dispersion in 
the local calibrating sample using the slope from (3). Steps
(2) through (4) are repeated until the slope and zero point converge.
Convergence is usually achieved within three or four iterations.
We obtain the following I--band TF relation:

\begin{equation}
I^c_T = -10.00 (\pm 0.08) (\log W^c_{50} - 2.5) - 21.32 
\label{eqn:ibandtf}
\end{equation}

For reference, the I--band--50\% TF relation derived using only the calibrators
(equivalent to Equation 9) is: $I_T^C = -9.87 (\log W^C_{50} - 2.5) - 21.30$,
which is virtually the same as Equation 11.
Finally, note that the internal extinction correction of T98 is expressed
in terms of 20\% linewidths.  The G97 database, on the other hand, lists 50\% linewidths.
We have made a compilation of galaxies with both published 20\% and 50\% linewidths
by examining the G97 survey and an H--band survey by Aaronson, Huchra, Mould and collaborators
(cf: Aaronson et al. 1982).
The mean ratio for 73 such galaxies is $W_{20} = 1.1 (\pm 0.03) W_{50}$.
When estimating the internal extinction, we calculate the corresponding 20\% 
linewidth using this ratio for all the cluster galaxy data.
In addition, the T98 internal extinction expressions are given as
a function of turbulence--corrected linewidth.
Thus, only in estimating the internal extinction, we apply this turbulence
correction to the linewidths following Tully \& Fouqu$\acute{\mbox{e}}$ (1985).

\subsection{Uncertainties in the Tully--Fisher Parameters}

To better understand the errors in the derived TF relations, 
we generated  a sample of 5000 random points
which followed a Tully--Fisher relation of dispersion 0.43 mag.  From this 
sample, 500 random subsamples of $N$ galaxies were drawn,
and the zero point and slope were calculated for each trial run.
We examined three cases: N=25, 500 and 2000.
For $N=25$ trial runs, the range of slope distribution is $\sim$5 times larger
than other two larger samples, suggesting that the sample solely consisting of
the $\sim$20 Local calibrators could give a slope estimate that is 
significantly different from the true value.
The cluster database used to estimate the TF slope was comprised of more than 500 galaxies
(for the I--band survey). Even for this larger sample, the slope distribution
has a 1$\sigma$ $RMS$ scatter of 0.09 mag.

The distributions of zero points obtained for the same set of 
simulations were also examined.  Again as expected, the smallest 
sample, in which 25 galaxies were picked during each trial run, has 
the largest standard deviation in the zero point distribution of 0.13 mag,
which in fact agrees with the actual zero point error calculated in
Equations 6--10.

We have determined the slope using a sample of $\sim$500 galaxies 
and then the zero point using a sample of $\sim$20 galaxies.
Thus, we formally adopt the uncertainties for the slope and zero point
of 0.09 and 0.13 mag respectively.
In Section 5.5, the sensitivity of the value of H$_0$ to the slope
uncertainty of the TF calibration will be investigated further.

\subsection{Dispersion}

The dispersion of the TF relation has three primary origins:
(1) intrinsic to the relation itself, due to departures from the 
``perfect''  physical correlations, such as 
Freeman's exponential disk law, or variations in mass--to--light ratio,
(2) observational errors in magnitude and linewidths, and
(3) for the case of a cluster TF relation, the physical depth of the cluster
itself or, for the case of the calibrators, the uncertainties in the distance
estimates of the individual galaxies.

The subject of the TF dispersion has long been of interest, as its value
can pose strong constraints on scenarios for galaxy formation
and evolution.
For example, Eisenstein and Loeb (1996) suggested that a dispersion exceeding
0.3 mag is expected for most galaxy formation histories.
Silk (1996) speculated that a tight TF relation could suggest an
unexpectedly late epoch of galaxy formation.
It is also important to understand the intrinsic scatter of the TF
relation, because it is directly related to the bias in the distance ladder
itself.   

Surprisingly, estimates of dispersion have been very ambiguous,
ranging from $\sim 0.1$ mag (e.g. Bernstein et al. 1994, Freedman 1990),
up to $\sim0.7$ mag (e.g. Kraan--Korteweg et al. 1988, Sandage 1994, Tammann 1999),
with many estimates around $0.2-0.3$ mag (e.g. Pierce \& Tully 1988,
Willick 1991, Courteau 1992, Bothun \& Mould 1987, Schommer et al. 1993).
More recent studies have shown furthermore that the TF dispersion is
a more complex function of galaxy types and sizes.
Using his LCO/Palomar Survey, Willick (1999) found that the dispersion
is as much as $\sim$0.75 mag for low--luminosity, low--surface brightness
galaxies, while for the higher luminosity, higher surface brightness
galaxies, it is $\leq$ 0.35 mag.
Giovanelli (1997) reported that the dispersion of the TF
relation is in fact a function of linewidth; it decreases from
fast--rotators ($\sigma \sim 0.2$ mag for $\log W \sim 2.6$)
to slow--rotators ($\geq 0.3$ mag for $\log W < 2.2$).
Readers are referred to Giovanelli (1997) and
Willick (1999) for an extensive discussion on this subject.
Here, we have a unique set of data:
the local calibrator data provide a sample in which the
distance to each galaxy is known with 10--15\% accuracy unlike  the
case of the cluster TF relation, where the cluster depth can be an additional
unknown parameter.
Freedman (1990) also used galaxies with accurately measured
Cepheid distances, but this sample now has 21 galaxies, instead
of five used by Freedman (1990).
Dispersions for BVRIH Tully--Fisher relations obtained from
our Cepheid calibrator galaxies were quoted in Section 3.1.
These are, however,  observed dispersions and are a combination
of observational errors and the intrinsic scatter of the TF relation itself,
added in quadrature.
We determined the intrinsic scatter of the TF relation for
each wavelength, by estimating the observational error for each calibrator
galaxy following Equation~10 of Giovanelli et al. (1997b).
We have also assumed that the dispersion is a constant
over the linewidth range $2.4 \leq \log W \leq 2.8$ spanned by the calibrators.
A typical observational error for galaxies is $\sim 0.30 \pm 0.05$ mag,
(whose main sources are photometric, distance modulus
uncertainty and the linewidth measurement uncertainty multiplied by the TF
relation slope,)
indicating that the intrinsic dispersions are $0.25, 0.22, 0.25, 0.20$ and $0.19$ mag
for B, V, R, I, and H$_{-0.5}$ respectively.
We have tested to see if this dispersion is a function of some observational
parameter, such as the color.  In Figure 2, we plot TF residuals as a function 
of color, which show no dependence, suggesting that
color is certainly not a candidate for the second parameter in TF
relations.

Another question to be addressed is whether a dispersion of 0.20--0.25 mag
is a value that could have been measured accidentally.  
In the previous section, the slope and zero point of the TF relation were
examined using a randomly created sample of 5000 galaxies.
A similar exercise is carried out here, except a sample of 5000 points
were generated assuming that the dispersion of the TF relation is 0.7 mag.
A random subsample of galaxies was again withdrawn  500 times. 
The dispersion of 0.2 mag is highly unlikely, with a probability
of practically zero.
The B-- and V--band TF relation dispersions were slightly higher than that of
other wavelengths, measured at 0.25 mag.
The chance of `accidentally' observing even this dispersion is remote, less than 1\%,
if the true dispersion is 0.7 mag.

Furthermore, when the cluster data, consisting of few hundred galaxies,
are put on the same TF plane,
one finds that its dispersion is less than 0.3 mag.
This confirms that the intrinsic dispersion of TF relation 
is relatively small.

\subsection{Color Dependence of Tully--Fisher Relations}

If the internal extinction correction were perfect, there should be no
dependence of the corrected color of the TF galaxies on inclination.
In Figure 3 (top), we examine the $(B-I)$ color
of the local calibrators as a function of inclination.
The solid circles represent the data corrected using Equation 1,
while the open circles indicate the uncorrected data.
There is a slight dip in this color--inclination relation around 
the inclination angle of 50$^{\circ}$, however, there is no overall
trend that is observable.
Also shown in the middle and bottom plots of 
Figure 3 are the $(B-I)$ vs inclination relations for the same
set of Local calibrator galaxies, but here, Han's 
(Equation A1 in Appendix A) and Giovanelli et al.'s (Equations A2 and A3) 
extinction corrections, respectively, 
were used.  
Although for both Han's and Giovanelli et al.'s corrections there may be
a slight decline in the relation for the large--inclination region,
this apparent trend could be partly due to small--number statistics.
One significant question is whether the value of H$_0$ is affected by the 
uncertainty  arising from the internal extinction corrections.
This will be discussed in detail in Section 5.7.

\subsection{Barred Galaxies in the Tully--Fisher Relations}

Among the 21 calibrator galaxies in our Tully--Fisher sample, seven
are barred galaxies.  In this section, we explore whether these
barred galaxies are in fact suitable TF calibrators or not.
On the left side of Figure 4, we have re--plotted B, I and H band TF relations
using the 20\% linewidth.
Seven barred galaxies, NGC~925, NGC~1365, NGC~3319, NGC~3351,
NGC~3627, NGC~4535, and NGC~4725 are shown with open circles.
The solid line in each plot represents the TF fit to all the
galaxies.
%For the B--band data, only six barred galaxies are
%shown as the photometric data for NGC~1365 is not available.
A striking feature is that the majority of them lie above the
TF relations, suggesting that they are systematically brighter
for a given rotational velocity.
This is true in all three wavelengths. 
(Although not shown here, the same applies to the V-- and R--band
TF relations).
In Table B1, a summary of published photometric and kinematical
inclination angles are listed. 
We use the kinematical inclination angles for the seven barred galaxies
and plot the TF relation on the right side.
The solid line in each plot shows the fit through all the
galaxies, using the kinematical inclination angles for the barred
galaxies. The dotted line is the fit through the TF relation
using the photometric inclination angles for all galaxies
(same as the solid line in the left plot).
The change in zero point in the I--band, for example, is 0.04 mag.
Also, for the I--band, the observed dispersion decreased considerably, 
from 0.37 mag down to 0.32 mag. For the B--band relation, it changed significantly
as well from 0.45 to 0.41 mag.
The dispersion remains unchanged when kinematical inclination angles
are substituted for all other galaxies, indicating that the
photometric determination of inclination angles for barred
galaxies are systematically different from the kinematical estimates.

Two possible explanations for the significant
difference in the measured photometric and kinematical
inclination angles of barred galaxies.
The first is that the star formation rate is higher
in the barred galaxies, although observations suggest that
the effect of a bar on the disk star formation rate
is unimportant (Kennicutt 1999).
In Figure 5, colors of the calibrator galaxies are shown
as a function of linewidth.  The barred galaxies are indicated
by open circles.  
In the plots on the left side, the photometric inclination angles
were used to derive the corrected magnitudes, while on the right side,
the magnitudes were corrected using the kinematical inclination angles
for seven barred galaxies.
Although most barred galaxies lie on the blue edge
of this correlation if the photometric inclination angles are used
(with one exception, NGC~3351, which is located at the red edge),
when the kinematical inclination angles are substituted for barred galaxies,
there are no longer any obvious differences between the distributions of
the barred and non--barred galaxies.
If an additional star formation is actually triggered by
the presence of a bar, then one would expect the deviation of
barred galaxies from the TF relation would be higher
for shorter wavelength.
We do not see such evidence, although our sample is rather small.
We find the deviations from the 
mean TF fit for barred galaxies are the same for all
five wavelengths.
The second explanation, which is more plausible, is that 
the inclination angles for barred galaxies are over--estimated 
if the photometric inclination angles are used.  
This would make the barred galaxies appear brighter for
a given linewidth, and would explain that adopting the kinematical
inclination angles would shift these galaxies closer to the TF relation.

As stated in Section 2.2.2,
in the following analysis in this paper, we will not use the kinematical
inclination angles, because
those for the I--band cluster survey galaxies
were derived strictly from the photometric observations.
In Figure 6, we show a TF relation for cluster galaxies.
Those classified as barred are shown by solid circles.
There is no obvious systematic offset between the barred and
non--barred samples here, although one would expect that
both the morphology identification and inclination estimates
might be more difficult for distant galaxies.
We have also examined the distribution of the ratio of number
of barred to non-barred galaxies as a function of linewidth or inclination,
but observed no significant difference between the cluster and calibrator
samples.

\section{Application of the Tully--Fisher Relation to Distant Cluster 
Samples and the Value of the Hubble Constant}
\label{section:tfapplication}

There have been numerous Tully--Fisher 
surveys of clusters of galaxies in various wavelengths.
Most recently, Giovanelli and collaborators presented a survey
of I--band photometry and radio linewdiths of 
$\sim$2000 spiral galaxies in clusters out to $\sim$10,000 km s$^{-1}$.
Aaronson, Huchra, Mould and collaborators
presented H--band aperture magnitudes and 20\% linewidths of galaxies
in clusters out to $\sim$15,000 km s$^{-1}$ (Aaronson et al. 1980, 1982, 1986
: hereafter AHM).
Bothun et al. (1985) presented a catalog of $UBVR$ multi-aperture photometry,
H--band photometry and 21cm HI observations for several hundred spiral galaxies
in 10 clusters.
Han and Mould (1990) published I--band magnitudes for galaxies in the AHM
survey.  
In this paper, we focus on the I--band survey by Giovanelli et al. (1997a),
as it is the most complete one to date, and presents total I--band
magnitudes measured from CCD observations, unlike in other surveys where 
aperture magnitudes were used.

\subsection{I--band Cluster Distances and the Hubble Constant}

The absolute calibration of the $BVRIH_{-0.5}$ Tully--Fisher relations was presented
in the previous section.  The I--band relation was also 
re--calibrated for specific application to the G97 database.  The second I--band
calibration can now be applied to the distant cluster surveys, which extend
out to $\sim$10,000 km s$^{-1}$, to determine H$_0$.

For a detailed description of the I--band database,  readers
are referred to Giovanelli et al. (1997a) who explain in detail
the selection criteria and the cluster galaxy membership.
Briefly, the G97 survey is comprised of clusters and groups of galaxies 
out to 9000 km s$^{-1}$.  The database consists of 50\%
linewidths, total $I-$band magnitudes which were extrapolated
to infinity (the largest isophotal ellipses measured 
were usually $\sim$24.5 mag arcsec$^{-2}$),
and photometric inclination angles measured from I--band images.

In Table 3, we summarize the clusters of galaxies that are used in
the following analyses to measure the Hubble constant.
Also listed in this Table are three velocities for each cluster:
velocities corrected for the Local Group frame, cosmic microwave
background frame (CMB), and corrected for the flow model which
will be described in Section 5.8. 

To be included in the TF analyses, 
the galaxies had to meet all of the following criteria:

(1) They do not deviate by more than twice the dispersion from the TF relation.
Three galaxies were excluded due to this: I1830 in the Fornax cluster, 
N2535 in Cancer, and N3832 in A1367.

(2) Galaxies must not be nearly face--on ($i \leq 40^{\circ}$)
because for these objects,  the inclination uncertainty and subsequently
the linewidth uncertainty is too large.
Also, galaxies should not be close to edge--on ($i \geq 80^{\circ}$)
since the extinction correction for these become unreasonably large.

(3) The linewidth distribution of galaxies in the clusters should be
similar to that of the calibrators.
In Figure 7, some examples of clusters are shown. 
In the same figure, the calibrator data are over-plotted on each cluster TF relation
with solid circles.
The local calibrators and the cluster samples do not all cover
the same rotational velocity range.  For some clusters, such as Eridanus,
Cancer, or Pegasus groups, the cluster sample extends to much smaller
linewidths. In order to make a consistent sampling between the two
sets of datasets, we apply a lower $\log W$ cutoff of 2.35.

(4) The internal extinction correction should not exceed 0.6 mag.
In Figure 8, the distributions of internal extinction values
for the cluster galaxies and calibrators are obtained using three separate
extinction corrections, that of T98, Han and Giovanelli et al.
(see Appendix A for the latter two corrections).
For the T98 and Giovanelli et al. corrections, 
there is a significant difference between the
two samples: the cluster galaxies seem to have a long upper--end tail
in their distribution of extinction values.
To make a consistent sampling between the calibrators and clusters,
we apply an upper cutoff of $A_{int,I} = 0.6$ mag.
It is also a matter of common sense to exclude galaxies more than
half of whose light is obscured ($A_{int} > 0.75$ mag).

The final $I-$band sample, referred to as the G97 sample, 
thus consists of 276 galaxies.

{\scriptsize
\tablenum{3}
\begin{deluxetable}{lcccccc}
\tablecolumns{7}
\tablewidth{0pc}
\tablecaption{\bf Clusters of Galaxies in the I--band Survey}
\tablehead{
\colhead{Cluster/} &
\colhead{R.A.\tablenotemark{a}} &
\colhead{DEC\tablenotemark{a}} &
\colhead{V$_{LG}$} &
\colhead{V$_{CMB}$} &
\colhead{V$_{Model}$} &
\colhead{N} \\
\colhead{Group} &
\colhead{hh mm ss} &
\colhead{dd mm ss} &
\colhead{km s$^{-1}$} &
\colhead{km s$^{-1}$} &
\colhead{km s$^{-1}$} &
\colhead{}
}
\startdata
 N383       & 01 04 30 & $+$32 12 00  & 5425 & 4924 & 5086 & 23 \nl
 N507       & 01 20 00 & $+$33 04 00  & 5350 & 5869 & 5016 & 14 \nl
 A262       & 01 49 50 & $+$35 54 40  & 5170 & 4730 & 4852 & 32  \nl
 A400       & 02 55 00 & $+$05 50 00  & 7258 & 7016 & 6983 &  26 \nl
 Eridanus   & 03 30 00 & $-$21 30 00  & 1641 & 1607 & 1627 &  35 \nl
 Fornax     & 03 36 34 & $-$35 36 42  & 1328 & 1380 & 1372 &  44 \nl
 Cancer     & 08 17 30 & $+$21 14 00  & 4652 & 4982 & 4942  &  29  \nl
 Antlia     & 10 27 45 & $-$35 04 11  & 2505 & 3106 & 2821 & 26 \nl
 Hydra      & 10 34 28 & $-$27 16 26  & 3454 & 4061 & 3881 & 27 \nl
 N3557      & 11 07 35 & $-$37 16 00  & 2702 & 3294 & 2957 & 12 \nl
 A1367      & 11 41 54 & $+$20 07 00  & 6316 & 6709 & 6845 & 38 \nl
 Ursa Major & 11 54 00 & $+$48 53 00  & 957 & 1088 & 1088 & 30  \nl
 Cen30      & 12 46 06 & $-$41 02 00  & 2756 & 3272 & 4445 & 39 \nl
 Cen45      & 12 47 57 & $-$40 22 23  & 4304 & 4820 & 4408 &  9 \nl
 Coma       & 12 57 24 & $+$28 15 00  & 6889 & 7143 & 7392 & 41 \nl
 ESO508     & 13 09 54 & $-$23 08 54  & 2662 & 3149 & 2896  & 17 \nl
 A3574      & 13 46 06 & $-$30 09 00  & 4308 & 4749 & 4617 & 20 \nl
 Pavo       & 20 13 00 & $-$71 00 00  & 3919 & 4027 & 4220 & 10 \nl
 MDL59      & 22 00 18 & $-$32 14 00  & 2621 & 2304 & 2664 & 26 \nl
 Pegasus    & 23 17 43 & $+$07 55 57  & 4109 & 3545 & 3874 & 19 \nl
 A2634      & 23 35 55 & $+$26 44 19  & 9516 & 8930 & 9142 & 27 \nl
\enddata
\tablenotetext{a}{Positions given in Epoch 1950.0.}
\end{deluxetable}
}

We measured the TF distances to all the clusters with 5 member galaxies or more
by applying Equation 11 to the I--band survey.
The results are shown in Tables 4 in which the cluster or group distances, $D$, 
and $V/D$ values (corresponding to the velocities with respect to the
Local Group, CMB and model) are listed for I--band clusters.
The uncertainty in the H$_0$ values is roughly a combination of observational
errors, errors in the TF zero points and slopes, and the dispersion of the TF relation.
This will be discussed in more detail in Section 6.

The top plot in Figure 9 shows the Hubble diagram for the I--band clusters.
Also in the bottom of the same figure, 
we plot the Hubble constant for each cluster as a function of velocity.
In order to measure the values of H$_0$, we use only those clusters 
satisfying (1) $V_{CMB} \geq 2000$ km s$^{-1}$, and 
(2) velocities with respect to the CMB and the flow field model are
within 10\% of each other.
This leaves us with a set of 15 clusters and groups of galaxies.
Also, it will be shown in Section 5.1.2 that there are no selection
biases for galaxies with $V_{CMB} \geq 2000$ km s$^{-1}$, thus
using those clusters with $V_{CMB} \geq 2000$ km s$^{-1}$ should be 
a reasonable choice.
Taking an average of 15 clusters,
we obtain H$_0$ = 73 $\pm$ 2 (random) $\pm$ 9 (systematic) using the I--band survey.
The random and systematic errors are discussed in detail in Section 6.

{\scriptsize
\tablenum{4}
\begin{deluxetable}{lcccccc}
\tablecolumns{7}
\tablewidth{0pc}
\tablecaption{\bf I--band Tully--Fisher Distances to Clusters}
\tablehead{
\colhead{Cluster} &
\colhead{N} &
\colhead{$m-M$} &
\colhead{D} &
\colhead{$V_{\mbox{LG}}/D$} &
\colhead{$V_{\mbox{CMB}}/D$} &
\colhead{$V_{\mbox{model}}/D$} \\
\colhead{} &
\colhead{} &
\colhead{(mag)} &
\colhead{(Mpc)} &
\colhead{(km/s/Mpc)} &
\colhead{(km/s/Mpc)} &
\colhead{(km/s/Mpc)}
}
\startdata
A1367       &  28 &  34.84 &  93.0 &  67.9  (11.0)  & 72.2  (11.7)  & 73.6  (11.9) \\
A2197       &   3 &  35.56 & 129.3 &  71.8  (11.4)  & 70.5  (11.1)  & 73.9  (11.7) \\
A262        &  22 &  34.18 &  68.7 &  75.3  (12.2)  & 68.9  (11.1)  & 74.2  (12.0) \\
A2634       &  20 &  35.36 & 118.0 &  80.6  (12.6)  & 75.7  (11.8)  & 77.5  (12.1) \\
A3574       &  14 &  34.03 &  64.1 &  67.2  (10.5)  & 74.1  (11.6)  & 72.0  (11.2) \\
A400        &  18 &  34.81 &  91.8 &  79.1  (12.3)  & 76.5  (11.9)  & 76.1  (11.8) \\
Antlia      &  16 &  33.30 &  45.6 &  54.9  ( 8.8)  & 68.1  (10.9)  & 61.8  ( 9.9) \\
Cancer      &  17 &  34.40 &  75.8 &  61.4  ( 9.8)  & 65.8  (10.5)  & 65.2  (10.4) \\
Cen30       &  22 &  33.25 &  44.7 &  61.7  (10.2)  & 73.2  (12.1)  & 99.5  (16.4) \\
Cen45       &   8 &  34.24 &  70.5 &  61.1  (10.0)  & 68.4  (11.2)  & 62.5  (10.3) \\
Coma        &  28 &  34.74 &  88.6 &  77.7  (12.2)  & 80.6  (12.6)  & 83.4  (13.0) \\
Eridanus    &  14 &  31.66 &  21.5 &  76.4  (12.4)  & 74.8  (12.1)  & 75.8  (12.3) \\
ESO508      &   9 &  33.07 &  41.1 &  64.8  (10.3)  & 76.7  (12.1)  & 70.5  (11.2) \\
Fornax      &  14 &  30.93 &  15.3 &  86.6  (14.0)  & 90.0  (14.5)  & 89.5  (14.4) \\
Hydra       &  17 &  33.90 &  60.2 &  57.4  ( 8.9)  & 67.4  (10.5)  & 64.5  (10.0) \\
MDL59       &  10 &  32.56 &  32.5 &  80.7  (12.6)  & 71.0  (11.1)  & 82.1  (12.8) \\
N3557       &   9 &  33.02 &  40.2 &  67.2  (11.1)  & 81.9  (13.5)  & 73.5  (12.1) \\
N383        &  15 &  34.19 &  68.9 &  78.7  (12.3)  & 71.5  (11.2)  & 77.3  (12.1) \\
N507        &   7 &  33.86 &  59.1 &  90.6  (14.0)  & 82.4  (12.8)  & 89.0  (13.8) \\
Pavo        &   5 &  32.71 &  34.8 & 112.6  (17.4)  &115.7  (17.9)  &121.3  (18.7) \\
Pavo2       &  13 &  33.62 &  52.9 &  81.3  (13.0)  & 83.2  (13.4)  & 87.9  (14.1) \\
Pegasus     &  14 &  33.73 &  55.8 &  73.6  (11.5)  & 63.5  ( 9.9)  & 69.4  (10.9) \\
Ursa-Major  &  16 &  31.58 &  20.7 &  46.3  ( 7.1)  & 52.6  ( 8.1)  & 52.6  ( 8.1)   
\enddata
\end{deluxetable}
}

\subsection{Comparison with Other Surveys}

As mentioned above, there are other surveys of clusters of galaxies that 
reach velocities $\sim$10,000 km s$^{-1}$, which are suitable
for the determination of H$_0$. 
The G97 survey was used as the primary survey in this paper because of
its completeness and quality.
In this section, we compare our I--band H$_0$ results with the distances
and values of H$_0$ obtained using other surveys.

\subsubsection{H--band}
The H--band cluster surveys come from two
sources: Aaronson et al. (1982) and Aaronson et al. (1986). 
We refer to the combined sample as AHM.  
In these surveys, rotational velocities were measured at the 20\% level,
and $H-$band aperture magnitudes were corrected to an isophotal aperture of
$\log A/D_1 = -0.5$ by adopting the system of the RC2 catalog (de Vaucouleurs,
de Vaucouleurs, and Corwin 1976).
These magnitudes are referred to as $H_{-0.5}$ (see Aaronson, Mould and Huchra
1980 for details of how the interpolation was made).
Also, in the original database of Aaronson et al. (1980), they
included a 3\hbox{${}^\circ$} additive term for inclination angles.
We are not adopting this convention, because we use the axis ratios
given by RC3 catalog.

Cluster membership in the \hband survey was not as well defined as for 
the \iband survey, where G97 thoroughly inspected
each cluster to determine which galaxies are likely members of the cluster.
We examined the distribution of galaxies in each cluster in both spatial and 
velocity space to identify any outliers. 
Several clusters, including A1367, Z74--23, Hercules and Pegasus, 
were split into two or three subgroups. %Detailed notes are found in Table 4.

The same four criteria listed above for the selection of the 
I--band survey galaxies are applied to the H--band AHM surveys.
Again, to make the cluster database consistent with the calibrators,
a lower cutoff of $\log W \geq 2.35$ is applied.
The final H--band sample thus consists of 163 galaxies in 26 clusters.

Following the method described in Section 3.2, the H--band TF relation
was derived using both the calibrators and the cluster data.
The resulting value of H$_0$ is 67 $\pm$ 3 (random) $\pm$ 10 (systematic).
The 10\% discrepancy between the H--band and I--band values of H$_0$ is
statistically significant, greater than the 2$\sigma$ level.
Tormen \& Burstein (1995) published a list of isophotal H magnitudes for
galaxies from the AHM survey, corrected using RC3
parameters.  
Substituting this set of data for both the calibrators and cluster
galaxies yields the same value of H$_0$;
use of these updated magnitudes does not reconcile
the disagreement between I--band and H--band H$_0$ values.
We defer further discussion of this discrepancy to Section 4.2.3.

\subsubsection{B-- and V--band Surveys}

Bothun et al. (1985: hereafter B85) presented a catalog of 21cm HI observations,
$UBVR$ multi-aperture photometry, and H--band photometry of several
hundred galaxies in 10 clusters.
The total $UBVR$ magnitudes were extracted from multiple--aperture
measurements by applying the aperture corrections estimated from the mean
growth curve derived from the RC2.
The average aperture correction amounted to 0.2--0.3 mag.
Of these 10 clusters, we selected five clusters, A1367, A400, A539,
Coma and Z74--23, which consisted
of at least five galaxies with good quality radio and optical data.
Using 57 galaxies with $\log W_{20} \geq 2.4$ in five clusters,
we obtain H$_0$ = 80 $\pm$ 8 km s$^{-1}$ Mpc$^{-1}$ (random) for B--band,
and H$_0$ = 73 $\pm$ 6 km s$^{-1}$ Mpc$^{-1}$ (random) for V--band.
However, we note that the H$_0$ value for A~539 is almost 3$\sigma$ larger
than those of other four clusters, in both B--band and V--band estimates.
Excluding this cluster, we obtain H$_0 = 72 \pm 2$ km s$^{-1}$ Mpc$^{-1}$
for B--band and 68 $\pm$ 2 km s$^{-1}$ Mpc$^{-1}$ for V--band.
These values lie between the I--band and H--band results.
We would put less weight on the B--band and V--band magnitudes as they
were extrapolated from aperture magnitudes, and our B-- and V--band calibrator 
magnitudes were not derived in the exactly the same manner.

\subsubsection{Discussion}

We have calculated H$_0$ using clusters of galaxies in four wavelengths:
B, V, I and H$_{-0.5}$.  
The derived values of H$_0$ vary from 67 up to 73 km s$^{-1}$ Mpc$^{-1}$.
In this section, we focus on the difference between the I--band and H--band results
which are based on more reliable databases, and are statistically better as 
the number of cluster
galaxies used to derive the value of H$_0$ is more than twice as large as those
in the B-- and V--band surveys.
It appears that the H--band TF relation yields consistently larger distances.
We examine the databases carefully with other available
published data in order to understand the root of the H$_0$ discrepancy.

{\bf Linewidth:}
The I--band analysis is based on a different linewidth set ($W_{50}$)
than the H--band analysis ($W_{20}$), so we first tested
whether the discrepancy in distances was caused by a subtle
error in one of the linewidth sets.
Because TF relations are steep, it is important to measure linewidths accurately.
Bothun et al. (1985) presented both 50\% and 20\% linewidths, and the 50\%
widths agree well with the published values of the G97 sample:
$<W_{\mbox{\small Bothun}}/W_{\mbox{\small G97}}> = 1.000 \pm 0.003$.
As noted in Section 2, we re-measured %on our own 
20\% and 50\% linewidths for the local calibrators,
and found that the $W_{20}/W_{50}$ ratios are consistent with those in the
cluster samples, $W_{20}/W_{50}$ = $1.07 \pm 0.03$ and $1.10 \pm 0.05$
respectively.
Furthermore, a list of galaxies that are in both G97 and AHM samples was
compiled.  The mean $W_{20}/W_{50}$ ratio of galaxies in the overlapping sample
($W_{20}/W_{50} = 1.10 \pm 0.08$)
again agrees well with the mean ratio of the local calibrators.
We have also compiled a set of $\sim$70 galaxies which had both
I--band and H--band magnitudes, and also 20\% and 50\% linewidths
measured, and confirmed that the derived values of H$_0$ are insensitive
to whether 20\% or 50\% linewidths are used.
Han \& Mould (1992) presented an I--band TF survey that used 
20\% linewidths. 
As an additional check we repeated the entire analysis using the Han \&
Mould database.  We obtained distances that are consistent with those
derived above using the G97 sample.
This leads us next to the various tests to confirm the photometric
quality of the cluster data and calibrators.

{\bf Photometry:}
We first tested the zero point of the G97 I--band magnitudes by
comparing them to those published by Han \& Mould (1992).
%, who  %presented the data for galaxies in the AHM survey.
There were 59 galaxies that overlapped in the two surveys:
they agree within 0.01 mag in the mean. 
On the other hand, Peletier and Willner (1993) published
$H_{-0.5}$ magnitudes for Ursa Major cluster galaxies.  We found that these also
agreed within 1$\sigma$ error with the AHM magnitudes.
The $R-I$ colors of the TF calibrator galaxies (whose magnitudes are given 
by Macri et al. 2000) agree well with those of normal spiral galaxies
presented in de Jong (1998).

Maeder et al. (priv. comm.) show that the concentric aperture photometry of
Aaronson et al is verified by recent 2MASS survey photometry.
But the consistency between calibrators' and clusters' isophotal
diameters, on which H$_{-0.5}$ magnitudes are based, is more
problematic. Diameters on the RC2 and RC3 systems differ by 0.1 dex
(Tormen \& Burstein 1995), and, while this apparently applies both to
calibrators and cluster galaxies, the absence of systematic differences
of the same order (equivalent to 0.2 mag in H$_{-0.5}$) cannot be guaranteed
without further work.

{\bf Color Distribution of Calibrators and Cluster Galaxies:}
The results above suggest that the difference in I--band and H--band
distances must reflect a systematic difference in measured $I-H_{-0.5}$
colors between the calibrator and cluster samples.
In the top panel of Figure 10, the I--band and H$_{-0.5}$--band TF distance moduli are compared.
In the middle plot, the $I - H_{-0.5}$ color of the same galaxies is shown as a function
of linewidth.
The calibrators are also shown by solid circles.
The most striking feature is that calibrators are redder compared
to the cluster sample galaxies (see below), which is the most likely reason
the H-- and I--band TF distances disagree at a 10\% level (top figure).
Environmental effects could possibly explain this difference.
However, as shown in the bottom plot of Figure 9, the B--I color distributions 
of the two sets do agree.
Furthermore, the R--I color distributions do not show any inconsistency between
the two sets.
K--S tests indicate that the chance that the $I-H_{-0.5}$
color distributions of the cluster galaxies and the calibrators are consistent 
is less than half of those  for other colors. 

{\bf Virgo and Ursa Major Clusters:}
Another test that indicates the preference for the I--band results
is to examine RIH$_{-0.5}$ TF distances to the Ursa--Major and Virgo clusters.
Pierce and Tully (1988) published $RI$ magnitudes of nearly 20 galaxies
in each of Ursa--Major and Virgo clusters.  Also $H_{-0.5}$ magnitudes
are available from the AHM survey.
We applied our calibrated $BRI$ TF relations to these samples to determine the
distances.
The $H_{-0.5}$ magnitudes for these galaxies were taken from the AHM survey.
For both clusters, we find that the $H_{-0.5}$ distances are systematically larger,
by 0.15 mag, or 7--8\% in linear distance.

From these comparisons, we tentatively conclude that the H--band TF distances
tend to be overestimated and should be used with caution:
that is, there is something systematically different about the $H_{-0.5}$
magnitudes of the nearby calibrators and distant cluster galaxies. 
The I--band distances are consistent with those of B--, V-- and R--band results,
and the H--band H$_0$ is inconsistent with other H$_0$ values by 2--4 $\sigma$.
Although they may be less affected by extinction, the $H_{-0.5}$ magnitudes are dependent
on isophotal diameter measurements and the concomitant aspect ratio corrections. 
There may be a systematic difference in $H_{-0.5}$ magnitudes of local
calibrators and cluster galaxies.  This will require further investigation,
but is beyond the scope of this paper, and the discrepancy is likely not
solved without a consistent set of total H--band magnitudes.
However, the measurement of total H magnitudes is made difficult by a sky background
which is more than 100 times brighter at H than I, and needs to be correctly subtracted
from the low surface brightness profiles in the outer parts of galaxies. Until better
photometry than the current H database becomes available,
the photometric systematics we have struck here may set the limits on H$_0$ 
measurements from the Tully-Fisher study.

\section{Discussion}

We have derived the value of H$_0$ by focusing on the survey by G97
who presented the 21cm linewidth and I--band photometry data for
several hundred galaxies in clusters out to $\sim$10,000 km s$^{-1}$.
In the previous section, this value was then compared with other surveys
of different wavelengths.
In this Section, we discuss how biases and systematic uncertainties
can affect the value of H$_0$, concentrating on the I--band database.

\subsection{Biases}

\subsubsection{Consistency between the Calibration and Distant Galaxies Samples}

Not only do the local calibrators and cluster sample galaxies have to be on
the same photometric and kinematical systems, they also need to be of
similar type.  
The $B-I$ color as a function of linewidth has often been used as this is a
good indicator of the variations in the recent star formation rate
($\sim$1 Gyr). It is also sensitive to the internal extinction
correction.  
Unfortunately for majority of the galaxies in the G97 I--band survey, there
are no $B$ magnitudes available. 
We thus compare our local calibrators with galaxies in the Virgo and
Ursa Major clusters whose photometric and linewidth data have
been published by Pierce \& Tully (1988).
These clusters should be representative of other clusters
of galaxies that are used in the two all--sky surveys we are calibrating,
especially since the spiral galaxies used are not located in the
immediate centers of the clusters, but in the ``infall'' regions.

In Figure 10 (bottom), we showed the $B-I$ color distributions of Virgo and
Ursa--Major cluster galaxies as a function of linewidth.
The local calibrators were over-plotted by solid circles.
The two cluster samples -- Virgo and Ursa Major -- show no
significant difference between their galaxy populations.
Furthermore, we see no significant difference between the
local calibrator sample and the two clusters.
This was not the case when Pierce \& Tully (1988) first examined this diagram;
they reported that six local calibrator galaxies from their sample were found
systematically on the blue edge of this color--linewidth relation.
The enlarged sample here indicates that there is no systematic difference 
between the local calibrators and cluster galaxy samples.

Another important issue that needs to be addressed is the difference
in environment between the calibration and cluster galaxies.
Of 21 calibrators, only seven are members of well defined groups/clusters of
galaxies: the Virgo, Fornax, and Leo~I clusters.  The rest are
members of loose groups or are field galaxies.
The application of the TF calibration based on such a sample to
the distant cluster data could potentially introduce another systematic
uncertainty, as cluster and field galaxies could be inherently different.
Using these seven calibrators only and fixing the slope to that obtained
from the cluster data, we derived an I--band TF zero point
(for a 50\% linewidth relation) of $-21.44 \pm 0.17$ mag.
This agrees within 1$\sigma$ of the zero point obtained in Equation~11,
suggesting that within our limited sample, there is no significant evidence
for environmental dependence in TF relations.
A more rigorous, statistically better defined analysis is presented by
Giovanelli (1997b) who examined the I--band TF residuals for galaxies grouped 
by the projected radial distance from the center of clusters. They found no zero 
point offset between subsamples.
In light of these results, we make no attempt to separate the field galaxies
from the cluster galaxies in the calibration.

\subsubsection{Selection Biases in the I--band}
\label{section:cpib}

In measuring the Hubble constant using distant cluster data,
it is important to account correctly for 
selection effects which include the Malmquist bias and 
cluster population incompleteness bias.
The former arises in field surveys, for example, where
galaxies are distributed uniformly and thus there are more galaxies per
unit volume at larger distances.  If each true distance is scattered
randomly, then more galaxies are scattered ``downward'' (i.e. smaller distances)
than ``upward''.  Thus, the distances are underestimated.
This type of bias was first examined in detail by Lynden--Bell et al. (1988).
Later, several authors expanded on this subject, including Gould (1993), 
Landy \& Szalay (1992) and Willick (1994).
%For our cluster samples, however, we do not need to be concerned with
%the Malmquist bias since the galaxies in the cluster are at the same 
%distance.
As for the cluster population incompleteness bias (CPIB), 
if not properly corrected for, the clusters would appear brighter
than they actually are because only the brightest galaxies in them are observed
in a flux--limited sample, leading to a biased value of the
Hubble constant.

There are numerous theoretical and empirical studies that have
dealt with the CPIB.
Teerikorpi, Bottinelli and collaborators introduced a quantitative
approach to this probelm (Teerikorpi 1984, 1987, Bottinelli et al. 1988).
Willick (1994) also presented a more extensive quantitative analysis of
the statistical bias.  
We will show in this section using the formalism presented by G97
that the CPIB in our sample is small, affecting the derived value of
the H$_0$ by less than 5\%.

Federspiel et al. (1994, hereafter FST94) investigated the bias properties
of the TF distances using a field sample of 1355 galaxies
from the Mathewson--Ford--Buchhorn survey.
One of the main conclusions presented by that study was that
in a flux--limited sample, at a given redshift (which is to first order,
distance), the Hubble constant is multi-valued for different values
of rotational velocity and apparent magnitude.
The trend is that for galaxies with similar rotational velocities,
the Hubble constant rises as the redshift becomes larger,
and for a given redshift, a higher H$_0$ is found for smaller galaxies
(smaller rotational velocities)
(for details, see Figures 1 and 2 of FST94).
They also showed that the zero point of the TF relation
is sensitive to the redshift, and to apparent magnitude as well.
Such biases clearly need to be corrected for if field samples are to 
be used to measure the Hubble constant.

We show, however, that for our cluster samples of the I--band spiral
galaxies, these biases are small.
Here, we examine the two distant cluster surveys which we are
calibrating to measure H$_0$, to see to what extent they may be affected
by the incompleteness biases discussed by FST94.

In Figure 11, \iband TF relations are shown for the G97
cluster sample.  Absolute magnitudes were calculated assuming $H_0 = 100$
km s$^{-1}$ Mpc$^{-1}$ and that velocities (in the CMB reference frame) give distances to 
first order.
The choice of H$_0$ here is not critical, as we are interested in examining
the {\it relative} offsets of different subsamples.
In the left panel of Figure 10, subgroups are divided by their
velocities, and the mean of each bin is shown.
With the exception of the nearest sample ($cz < 2000$ km s$^{-1}$),
there is no significant difference among the subsamples.
The slight offset in the zero point of the nearest sample is most likely
due to its complex velocity field not being characterized effectively by 
the CMB reference frame.
The right panel of Figure 10 shows subgroups divided by apparent 
magnitudes.  If the \iband TF sample is being severely affected by
its flux limit, then we should observe a systematic trend in the zero points
of the subsamples.  That is not the case, however.
The four subsamples, although some suffer from small number statistics,
show no significant dependence of TF zero point on the apparent magnitude.

Lack of selection biases suggests
that the incompleteness bias in the G97 \iband sample is
not severe, especially if the TF sample is restricted, for example
to those galaxies with larger rotational velocities ($\log W > 2.35$ -- 
see Figure 7).
In the next subsection (5.1.3), we will show quantitatively 
that cluster population incompleteness
bias for our cluster samples is at most $0.03 - 0.05$ mag, or 1--2\% in 
distance, and hence in the value of H$_0$.

\subsubsection{Cluster Population Incompleteness Bias}
\label{section:clusterpopbias}

We showed in the previous section that the selection bias with respect
to apparent magnitude or redshift is negligible.
In this section, we take a more quantitative approach to correct for 
selection biases.

There are several approaches to correct for CPIB.
Giovanelli et al. (1997b) presented a thorough, comprehensive method
to correct for this bias, which we apply to the \iband\ survey.
The basic idea is that the observed luminosity function for each cluster
is compared with the expected one derived from a theoretical luminosity
function such as the Schechter function.
Then a random sample is drawn from the model to 
represent the observed cluster.
After 1000 iterations, the zero point of the TF relation obtained from 
the simulated sample is compared to that of the observed one, to estimate the bias.
The detailed description of this CPIB method is found in Giovanelli 
et al. (1997b).

The cluster population incompleteness bias amounts to, on the average,
$\sim 0.03 - 0.05$ mag, corresponding to $\sim1 - 2$\% in distance.
Unlike what some authors have claimed in the past (e.g. Sandage, Tammann
and collaborators),  incompleteness bias is almost negligible.  
For example, Sandage (1994) had suggested that one of the reasons that the TF method
gave a high Hubble constant value was because only the brightest galaxies
in the cluster were observed: in particular, he believed that only the
``upper'' half of the TF relation was being considered.  This is not 
what is actually happening. These simulations show that the galaxies with
$\log W \geq 2.5$ are virtually unaffected by the cluster population incompleteness bias.
Thus the overall bias per cluster amounts to
only a couple of percent in distance. %%very well put !

\subsection{Systematic Uncertainties in H$_0$}

It is important to address how  systematic uncertainties
can affect H$_0$.
In this section, we focus on the following:
(1) the slope and shape of the TF relation, concentrating on 
how much a use of a wrong slope value can affect the H$_0$,
and also whether the application of quadratic fits would significantly
change the H$_0$ values;
(2) the metallicity dependence of the Cepheid period--luminosity
relation;
(3) the internal extinction correction method used; and
(4) velocity flow field models, focusing on how dependent the
value of H$_0$ is on the velocity reference frame adopted.
Table 5 summarizes all the values of H$_0$ derived for each test,
details of which are described below.

{\scriptsize
\tablenum{5}
\begin{deluxetable}{lc|lc}
\tablecolumns{2}
\tablewidth{0pc}
\tablecaption{\bf Systematic Errors affecting the I--band H$_0$}
\tablehead{
\colhead{Errors} &
\colhead{H$_0$} &
\colhead{Errors} &
\colhead{H$_0$} 
}
\startdata
 & & &  \nl
{\it TF Slope} &  & {\it Extinction Correction} &  \nl
$\:\:\:\:$$-11.00$ &  $71 \pm 2$ & $\:\:\:\:$T98               & $73 \pm 2$ \nl
$\:\:\:\:$$-10.00$  &  $73 \pm 2$ & $\:\:\:\:$Giovanelli et al. & $74 \pm 2$ \nl
$\:\:\:\:$$-9.00$  &  $75 \pm 2$ & $\:\:\:\:$Han               & $73 \pm 1$ \nl
 &  & &  \nl
Quadratic Fits &  $73 \pm 2$ & {\it Velocity Models} & \nl
 &  & $\:\:\:\:$Local Group  &  $71 \pm 3$ \nl
With Metallicity Correction    &  $70 \pm 2$ & $\:\:\:\:$CMB          &  $73 \pm 2$ \nl
Without Metallicity Correction &  $73 \pm 2$ & $\:\:\:\:$Infall Model &  $74 \pm 3$ \nl
 & & &  \nl
\enddata
\tablenotetext{*}{Unless otherwise indicated, the velocities in the
CMB reference frame were used.}
\end{deluxetable}

}

\subsubsection{Dependence of $H_0$ on TF Slope and Shape}

The effect of slope uncertainties on the value of H$_0$ is small.
We estimated the Hubble constant with 
three different assumptions for the I--band slope of $-9.00, -10.00$, and $-11.00$.
We obtain 75 $\pm$ 4, 73 $\pm$ 2, and 71 $\pm$ 2
km s$^{-1}$ Mpc$^{-1}$ respectively.
Errors quoted here include only the random uncertainties.
In this exercise, we changed the value of the slope by 1.0; 
in reality, this corresponds to more than 10$\sigma$, which suggests
that a use of incorrect TF slope will
affect H$_0$ insignificantly, perhaps 2-3$\%$ at most.

In Appendix C, quadratic TF relations were derived for each wavelength.
To demonstrate that the changes in the shape of the TF relation affects
H$_0$ very little, we applied the quadratic equation 
to the I--band cluster data.  
Note that the equation (C4) was derived for 20\% linewidth.
Here, we used a similar relation derived for the 50\% linewidth.
This yielded H$_0$ = 73 $\pm$ 2 km s$^{-1}$ Mpc$^{-1}$.

\subsubsection{Effects of Metallicity Dependence of the Cepheid PL Relation on $H_0$}

Whether the Cepheid period--luminosity relation depends on the metallicity
of the Cepheid variable stars has been a particularly important concern for the H$_0$
HST Key Project.  It may be one of the larger sources 
of systematic errors in the Cepheid distance scale.  
The metallicities of the Key Project galaxies vary almost by
an order of magnitude.
All the distances have been determined with
respect to that of the LMC, for which we adopt 
$(m-M)_0 = 18.50 \pm 0.13$ mag.  %0.15 is non-standard
%However, the metallicity of the LMC is $\log (O/H) + 12 = 8.50 \pm 0.15$ (ref).
However, the metal abundances of the TF calibrators cover
a wide range, up to five times higher than LMC.
Several observational and theoretical studies claim to
have constrained this metallicity effect (see Kennicutt et al. 1998).
Some estimates suggested a dependence of as much as 0.8 magnitude 
per factor of 10 increase in abundance (e.g. Gould 1994).
Because most of the galaxies used in our TF calibration are more
metal rich than the LMC, if such a steep dependence indeed existed, 
it could potentially modify the TF zero points significantly, and hence
the derived value of $H_0$.
%The change also works in a way to lower the value of H$_0$.

We have re--derived TF relations using the metallicity dependence
of $\gamma = \delta (m-M)_0 / \delta [O/H] = -0.24 \pm 0.16$ mag/dex, 
derived from the differential test of Cepheid distances to 
two fields in M~101 (Kennicutt et al. 1998).
Figure 12 shows the \iband\ TF relation
without corrections for the metallicity dependence of Cepheid PL relation
(solid circles) and with corrections (open circles).
The metallicity--corrected set of $BVRIH_{-0.5}$ TF relations are:

\begin{equation}
B^c_{T,Z} = -8.02 (\pm 0.72) (\log W^c_{20} - 2.5) - 19.78 (\pm 0.11)
\end{equation}
\begin{equation}
V^c_{T,Z} = -9.00 (\pm 0.83) (\log W^c_{20} - 2.5) - 20.34 (\pm 0.12)
\end{equation}
\begin{equation}
R^c_{T,Z} = -8.89 (\pm 0.72) (\log W^c_{20} - 2.5) - 20.67 (\pm 0.11)
\end{equation}
\begin{equation}
I^c_{T,Z} = -9.42 (\pm 0.76) (\log W^c_{20} - 2.5) - 21.15 (\pm 0.12)
\end{equation}
\begin{equation}
H^c_{-0.5,Z} = -11.21 (\pm 0.87) (\log W^c_{20} - 2.5) - 21.80 (\pm 0.14)
\end{equation}

We have also re--derived the I--band TF relation using the second
calibration method which was actually applied to the distant
clusters to derive H$_0$.  This method incorporated the cluster galaxy
data, using the 50\% linewidths:
\begin{equation}
I^c_{T,Z} = -10.00 (\pm 0.12) (\log W^c_{50} - 2.5) - 21.41
\end{equation}

The metallicity--corrected zero points above are brighter,
as most of the galaxies in this particular sample are
more metal rich than the LMC.  
Following the same exercise shown in Section 3.2, we have calculated
the TF zero point by adopting the slope measured from the cluster
galaxies.  The value of H$_0$ derived using the Cepheid distances
corrected for their metallicity dependence is H$_0$ = 70 $\pm$ 2
km s$^{-1}$ Mpc$^{-1}$, $\sim$ 4\% smaller than one obtained without
any corrections.

\subsubsection{Effects of the Uncertainties in the Adopted Internal
Extinction Corrections on $H_0$}

One of the largest corrections applied to the raw magnitudes is the internal
extinction correction, for which we adopted the expressions derived by T98.  
The internal extinction for highly inclined galaxies
amounts to more than 0.5 mag.
In Table~A1 (in Appendix A), 
we listed values of internal extinction for each Cepheid calibrator
galaxy determined via three separate methods: Tully et al. (1998),
Giovanelli et al. (1994) and Han (1992). The large dispersion between
these methods suggests that the internal extinction corrections are not
being determined uniquely.  Can the choice of extinction correction
method affect the measured value of $H_0$?

To quantify such an effect, we derived H$_0$ using all three above
methods of internal extinction correction.  
For each set of calculations, magnitudes of both calibrator and
cluster galaxies are treated using the same corrections for consistency, 
then the slope and zero point of the \iband TF
relation are estimated, as is 
the incompleteness bias for each cluster, and finally $H_0$ is calculated
from the measurements of TF distances to the clusters.
%The results are shown in Figure 14.  
For three extinction correction methods, T98, Giovanelli's and Han's, we obtain 
H$_0$ = 73 $\pm$ 2, 74 $\pm$ 2 and 73 $\pm$ 1 km s$^{-1}$ Mpc$^{-1}$ respectively. 
As we had expected, H$_0$ is virtually insensitive to the extinction correction
method used.  

Furthermore, there is of course the inclination uncertainty which
propagates through the extinction correction eventually to the
error in H$_0$.  
A full error analysis is provided in the next section.

\subsubsection{Dependence of $H_0$ on the Large--Scale Velocity Field}

One of the major remaining uncertainties in the determination of the H$_0$
is the correction of the observed velocities of galaxy clusters for
large--scale peculiar motions.  
It is now clear that there are motions on scales of tens of Mpc's
with amplitudes up to few hundreds of km s$^{-1}$ with respect to the CMB.
However, the exact nature of these motions is still unclear (Lauer and 
Postman 1994; Riess, Press and Kirshner 1995), especially in terms of
what precisely are the causes of these motions.
In Table 3, we listed three velocities: with respect to the
Local Group reference frame, CMB frame, and the modeled flow field,
all of which were used to derive H$_0$ for each cluster for
I--band surveys.% (Tables 5 and 6).
The observed heliocentric velocity ($V_H$) of the galaxies are 
corrected to the centroid of the
Local Group following Yahil, Tammann and Sandage (1977),
and also to the CMB frame via the relation from Giovanelli et al. (1998).
The flow field model adopted in this paper is a multi--attractor model which included
the Virgo cluster, the Great Attractor and the Shapley concentration.
The details of the flow field model are presented in Mould et al. (1999).

The flow field model is explained briefly as follows. 
It is a multi--attractor linear model based on Han and Mould (1990) and
Han (1992).  The model is comprised of three attractors: 
(1) the Virgo cluster at (RA,DEC,V$_H$) = (12.472$h$, +12\Deg67, 1035km s$^{-1}$),
with an assumed radius of 10 degrees;
(2) the Great Attractor at (13.333$h$, -44\Deg00, 4600km s$^{-1}$),
with an assumed radius of 10 degrees;
and (3) the Shapley concentration at (13.50$h$, -31\Deg0, 13800km s$^{-1}$),
with a radius of 12 degrees.
The flows toward each attractor are assumed to be independent of each other, thus the
total velocity correction is a vector sum of all three peculiar motions.

As shown in Table 5 in which we have listed H$_0$ estimates using three different
velocities, H$_0$ is not sensitive to the velocity field adopted.
The value of H$_0$ remains unchanged whether we adopt the velocities in the
CMB frame of reference or those derived using the infall model.

\bigskip
\bigskip
\bigskip

\section{Uncertainty in the Hubble Constant}

Understanding all the sources of errors, both systematic and random, and
how they propagate into the uncertainty in the value of the Hubble constant 
is almost as challenging as the actual measurement of H$_0$\ itself.
Table 6 shows how the uncertainty in H$_0$ is derived.
Here, we focus on the derivation of uncertainties for the I--band survey,
as that for the other surveys should be similar.

The first part describes the errors in the Cepheid distance scale.
Much of this has been discussed in detail in previous papers of this series.  
The largest systematic error originates in the uncertainty of the distance
to LMC, with respect to which all of the Cepheid distances have been measured.
Combining the LMC distance uncertainty with other errors such as those in 
the HST photometric zero points, extinction treatment and in the photometry
of Cepheid variables themselves, the total random and systematic errors
of the Cepheid distance scale add up to 0.08 and 0.16 mag respectively.

Next, the uncertainties in the TF distance scale are considered.
Using simulations, we have determined that the errors in the zero points
and slopes of the TF relations are 0.09 and 0.13 mag respectively.
For the H--band TF relation, however, we assign a larger error for its
zero point (0.25 mag) because of the systematic uncertainties in the photometry of
galaxies between the calibrator and cluster samples.
This was discussed in Section 4.2.
A typical error in the measured apparent magnitude of cluster galaxies
is $\sim$0.04 mag.  We have also assigned a typical error of 15 km s$^{-1}$
to the linewidth measurements, which is a conservative estimate.  
For a typical galaxy in our sample with $\log W \simeq 2.6$, a 15 km s$^{-1}$
error is equivalent to 0.04 mag.
There is in addition an uncertainty of 0.04 mag in the internal extinction correction,
which was estimated from comparing three different correction methods (Appendix A).  
Furthermore, the Galactic reddening adds an uncertainty of 0.02 mag.
These uncertainties, however, are reduced by the number of target galaxies
in each cluster, $\sqrt{n}$.  Thus the total random error in the TF magnitudes
per cluster is only 0.03 mag, assuming a typical number of galaxies
per cluster of 16.  The total linewidth error is 0.01 mag, again assuming
$n$ = 16.

{\scriptsize
\tablenum{6}
\begin{deluxetable}{llll}
\tablecolumns{4}
\tablewidth{0pc}
\tablecaption{\bf Error Budget (I--band)}
\tablehead{
\colhead{} &
\colhead{Source} &
\colhead{Error\tablenotemark{a}} &
\colhead{Notes} \nl
}
\startdata
1.  &  {\bf Cepheid Distance Scale } & & \nl
  &  {\it A. LMC True Modulus}    & $\pm$0.13  & Adopted from Westerlund (1997) \nl
  &  {\it B. LMC PL Zero Point} & $\pm$0.02  & From Madore \& Freedman (1991) \nl
S1.1 &  LMC PL Systematic Error   & $\pm$0.13  &  A and B added in quadrature \nl
   &  {\it D. HST V--Band Zero Point}\tablenotemark{b} & $\pm$0.03\tablenotemark{b} & \nl
   &  {\it E. HST I--Band Zero Point}\tablenotemark{b} & $\pm$0.03\tablenotemark{b} & \nl
S1.2 & Systematic Error in the Photometry & $\pm$0.09 & $\sqrt{D^2(1-R)^2 + E^2R^2}$, $R=A(V)/E(V-I)=2.45$ \nl
R1.1 & Random Error in the Photometry & $\pm$0.05 & From DoPHOT/ALLFRAME comparison \nl
   & {\it F. $R_V$ Differences Between Galaxy and LMC} & $\pm$0.014 & See Ferrarese et al. 1998 for details \nl
   & {\it G. Errors in the adopted value for $R_V$} & $\pm$0.01 & See Ferrarese et al. 1998 for details \nl
R1.2 & Random Error in the Extinction Treatment & $\pm$0.02 & F and G added in quadrature \nl
   & {\it H. PL Fit (V)}  &  $\pm$0.05\tablenotemark{c} &  \nl
   & {\it I. PL Fit (I)} & $\pm$0.04\tablenotemark{c} &  \nl
R1.3 & Random Error in the Cepheid True Modulus\tablenotemark{d} & $\pm$0.06\tablenotemark{c} & H and I partially correlated \nl
R$_{PL}$ & Total Random Error & {\bf $\pm$0.08} & R1.1, R1.2 and R1.3 added in quadrature \nl
S$_{PL}$ & Total Systematic Error & {\bf $\pm$0.16} & S1.1 and S1.2 added in quadrature \nl
  &   &  &  \nl
2. & {\bf Tully--Fisher Distance Scale} & & \nl
  & {\it A. Error in the measured apparent magnitude} & $\pm0.10\tablenotemark{c}/\sqrt{n}$\tablenotemark{f} & \nl
 &  {\it B. Error in the corrected linewidth} & $\pm0.04$\tablenotemark{c}$/\sqrt{n}$ & \nl
 & {\it C. Error in the internal extinction correction} & $\pm0.04\tablenotemark{c}/\sqrt{n}$ & average value from Table A1 \nl
 & {\it D. Error in the Galactic reddening} & $\pm0.02$ & Typical value given in \S6 \nl
R2.1 & Random Error in the TF Magnitudes & $\pm0.03$ & A,C and D added in quadrature with $n = 16$ \nl
R2.2 & Random Error in the linewidth & $\pm$0.01 & $n$ assumed to be 16 \nl
R2.3 & Intrinsic Dispersion of TF Relations & $\pm$0.20/$\sqrt{n}$ & \nl
R2.4 & Random Error on the TF moduli & $\pm0.12$ & $\sqrt{R2.1^2+(10.0 \times R2.2)^2 + R2.3^2}$ \nl
S2.1 & Systematic Error on the TF Zero Point & $\pm0.13$ & from Equation 9 \nl
S2.2 & Systematic Error in the internal extinction correction & $\pm0.04$ & from Table A1 multiplied by $\Delta A_I$ (cluster - calibrators) \nl
S2.3 & Systematic Error due to the metallicity dependence of & & \nl
   & $\:\:\:\:$ Cepheid distance scale & $\pm0.08$ & see Section 5.2.2 \nl
R$_{TF}$ & Total Random Error on the TF Distance Moduli & {\bf $\pm$0.12} & R2.4 \nl
S$_{TF}$ & Total Systematic Error on the TF Distance Moduli & {\bf $\pm$0.22} & S2.1, S2.2, S2.3, and S$_{PL}$ added in quadrature \nl
  &   &  &  \nl
3. & {\bf Hubble Constant} & & \nl
R3.1 & Random Error on the TF Cluster Velocities & $\pm300$\tablenotemark{e} & in km/sec \nl
%R3.2 & Random Error due to the variation of H$_0$ obtained  &  & \nl
%   &    $\:\:\:\:\:\:\:\:$ for  BVH$_{-0.5}$ surveys & $\pm$ 10 & in km/sec/Mpc \nl
R$_{H_0}$ & Total Random Error on H$_0$ & {\bf $\pm$ 2} & $\sqrt{(R3.1/v)^2 + (0.46 H_0 R_{TF})^2}/\sqrt{N}$, in km/sec\tablenotemark{g} \nl
S$_{H_0}$ & Total Systematic Error on H$_0$ & {\bf $\pm$ 7} & $0.46 H_0 S_{TF}$ in km/sec/Mpc \nl
\enddata
\tablenotetext{a}{Errors are in magnitudes for sections 1 and 2, and as indicated for section 3.}
\tablenotetext{b}{Contributing uncertainties from the Holtzman et al. (1995) zero points, and the
long--versus short uncertainty, combined in quadrature.}
\tablenotetext{c}{The values quoted are typical but individual cases vary slightly.}
\tablenotetext{d}{The partially correlated nature of the derived PL width uncertainties
is taken into account by the (correlated) de--reddening procedure, coupled with the largely
'degenrate--with--reddening' positioning of individual Cepheids within the instability strip.}
\tablenotetext{e}{Giovanelli's analysis of the velocity field.}
\tablenotetext{f}{Each cluster is assumed to have $n$ TF galaxies.}
\tablenotetext{g}{$v = 5000$km/s, and $N$ clusters of galaxies are used to derive the H$_0$.}

\end{deluxetable}
}

We learned in Section 3.4 that the intrinsic dispersion of the TF relation
is $\sim$ 0.20--0.25 mag.  Again, the uncertainty in the TF cluster distance due
to the intrinsic dispersion is reduced by a factor $\sqrt{n}$.  
The total random error in the TF distance moduli is therefore
the photometric error, the linewidth error multiplied by
the slope, and the intrinsic dispersion combined in quadrature.
Note that we have excluded the error in the TF slope.
We have shown in Section 5.2.1 that the slope uncertainty has a negligible
effect on the value of H$_0$.

The total systematic error in the TF distance modulus consists of three
terms.  The first is the error in the TF zero point calibration, which we
estimated to be 0.09 mag using Monte--Carlo simulations.
Another term is the systematic uncertainty arising from the internal
extinction correction.  
We have excluded any galaxies that required an internal
extinction correction of 0.6 mag or more, however, there still might be
a slight systematic error due to the fact that the {\it mean}
extinction corrections are not the same for the clusters and calibrators.
We have included this uncertainty (0.04 mag) in the total systematic
error of the TF distance moduli.
Adding these two terms in quadrature with the systematic error of the
Cepheid distance scale (0.16 mag), we estimate the total systematic
error of the TF distance scale to be 0.22 mag.

Finally, the uncertainty in the value of H$_0$\ can be determined.
In addition to the total random error on the TF distance
moduli derived above, the random error of the cluster velocities needs
to be considered.  In the last section, it was shown that the value of
H$_0$ was affected by the velocity flow field adopted, due to 
uncertainties arising from not understanding fully what the sources
of peculiar motions of clusters are, and also due to the non--uniform
distribution of clusters with respect to the CMB dipole.
Giovanelli et al. (1998) report that the distribution of peculiar
velocities of clusters in the CMB reference frame has a dispersion of 
300 km s$^{-1}$.  We assign this 300 km s$^{-1}$ as an additional
random uncertainty in the determination of the H$_0$ value.

Combining all the errors, we obtain H$_0 = 73 \pm 2 \pm 7$ in the I--band
(the first and second uncertainties correspond to random and systematic
respectively), $72 \pm 2 \pm 11$ in the B--band, $68 \pm 2 \pm 10$ in the V--band,
and $67 \pm 2 \pm 10$ in the H--band.
Again, larger systematic errors for the H--band result are due to
its less certain TF relation zero point.
We have also adopted larger systematic errors for the B-- and V--band
results as for these, the photometric data for calibrator and cluster
galaxies were not derived in a consistent manner, although both use
total magnitudes.
We take the weighted average of these four estimates to derive
the value of the Hubble constant, where the weights are defined as the
inverse square of the quadrature sum of the errors that are intrinsic
to the TF relations.  These include the random error on the TF moduli
(R2.4 in Table 6), the systematic error on the TF zero point (S2.1),
and the systematic error in the internal extinction correction (S2.2).
In Table 6, the error estimates for the I--band TF relation were listed.
For other wavelengths, the extinction correction uncertainties are
scaled by 2.75 for B--band, 1.5 for V--band and 0.5 for H--band.
Thus, the total random error on the TF moduli are 0.12 mag, 0.11 mag,
0.12 mag and 0.12 mag for B, V, I and H--band TF relations respectively.
As indicated above, the systematic error on the TF zero point is
0.13 mag for the I--band, while for others, we have
adopted 0.25 mag.
The weighted value of the Hubble constant is therefore:
\begin{equation}
H_0 = 71 \pm 4 ~\mbox{(random)} \pm 7 ~\mbox{(systematic)} ~\mbox{km s}^{-1} 
~\mbox{Mpc}^{-1}.
\end{equation}

\section{Summary}
The HST Key Project on Extragalactic Distance Scale has contributed
enormously in building the cosmic distance scale.
Taking the Tully--Fisher relation alone as an example, the number
of its calibrators have quadrupled as a result of Cepheid observations
using the HST.
This significantly improved the accuracy of the TF relation zero point,
and effectively the value of H$_0$.
Furthermore, numerous all--sky surveys, such as the one we used in
this paper by Giovanelli et al., allowed us to extend the galaxy distances
far out enough such that their peculiar velocities are significantly small
compared to their recession velocities.

{\scriptsize
\tablenum{7}
\begin{deluxetable}{clll}
\tablecolumns{4}
\tablewidth{0pc}
\tablecaption{\bf Estimates of H$_0$ using Tully--Fisher Relation}
\tablehead{
\colhead{H$_0$} &
\colhead{Method\tablenotemark{a}} &
\colhead{Sample} &
\colhead{Source} 
}
\startdata
$50 \pm 4$ & BTF & Virgo & Sandage \& Tammann 1976 \nl
$80 \pm 8$ & BTF & Virgo & Tully \& Fisher, 1977 \nl
$76 \pm 8$ & BTF & Virgo & Bottinelli \& Gouguenheim 1977 \nl
$61 \pm 4$ & IRTF & Virgo and Ursa Major & Aaronson, Mould \& Huchra, 1979 \nl
$65 \pm 4$ & IRTF & Virgo &  Mould, Aaronson \& Huchra, 1980 \nl
$95 \pm 4$ & IRTF & 4 clusters $<$6000 km/s & Aaronson et al. 1980 \nl
$63 \pm 10$ & $(H-I) vs \log W$ relation & Virgo & Tully, Mould \& Aaronson 1982 \nl
$71 \pm 2$ & VTF,RTF,IVTF  & Clusters $<$6,000 km/s  & Visvanathan, 1982 \nl
$82 \pm 10$ & BTF,IRTF & Groups $<$6000 km/s  &  Aaronson \& Mould, 1983 \nl
$74 \pm 11$ & VTF,RTF,IVTF & Clusters $<$6000 km/s & Visvanathan, 1983 \nl
$55 \pm 9$ & IRTF & Coma tied to Virgo &  Sandage \& Tammann, 1984 \nl
$91 \pm 3$ & IRTF & 20 Sc I's $<$13,500 km/s &  Bothun et al. 1984 \nl
$92 \pm 1$ & IRTF & Clusters $<$11,000 km/s &  Aaronson et al. 1986 \nl
$85 \pm 10$ & B,R,IRTF & Virgo and Ursa Major  &  Pierce \& Tully 1988 \nl
$57 \pm 1$ & BTF, IRTF & Virgo &   Kraan-Korteweg, Cameron \& Tammann, 1988 \nl
$68 \pm 8$ & BTF & Virgo & Fouque et al. 1990 \nl
$92^{+21}_{-17}$ & BTF & Coma &  Fukugita et al. 1991 \nl
$73 \pm 4$ & IRTF & Clusters $<$11,000 km/s &  Mould et al. 1996 \nl
$69 \pm 8$ & IRTF & Virgo, Ursa--Major & Shanks 1997 \nl
$82 \pm 10$ &  BTF  & Virgo &  Yasuda, Fukugita \& Okamura, 1997 \nl
$69 \pm 5$  &  ITF  &  Clusters $<\sim$10,000 km/s & Giovanelli et al. 1997a \nl
$53 \pm 5$ & BTF magnitudes & KLUN\tablenotemark{b} &  Theureau et al. 1997 \nl
$57 \pm 5$ & BTF diameters  & KLUN\tablenotemark{b} &  Theureau et al. 1997 \nl
$57 \pm 7$ & BTF & Clusters $<$11,000 km/s tied to Virgo & Federspiel, Tammann \& Sandage 1998 \nl
$56 \pm 3$ & BTF  & KLUN (HST Calibration) &   Theureau 1998 \nl
$51 \pm 4$ & BTF  &  KLUN (HIPPARCOS calibration) &   Theureau 1998 \nl
$53 \pm 3$ & BTF & Field galaxies $<$5000 km/s & Tammann 1999 \nl
$56 \pm 3$ & BTF & Clusters $<$11,000 km/s tied to Virgo  & Tammann 1999 \nl
$77 \pm 4$ & ITF & Clusters $<$8,000 km/s & Tully 1999 \nl
$76 \pm 3$ & ITF & Clusters $<$8000 km/s &  Madore et al. 1999 \nl
$52 \pm 5$ & BTF inverse diameter & KLUN & Ekholm et al. 1999 \nl
$53 \pm 6$ & BTF inverse magnitude & KLUN & Ekholm et al. 1999 
\enddata
\tablenotetext{a}{ITF: I--band TF; IRTF: IR TF; BTF: B--band TF;  VTF: V--band TF;
RTF: R--band TF; IVTF: 10500\AA}
\tablenotetext{b}{KLUN stands for Kinematics of the Local Universe, which consists
of a sample of 5171 spiral  galaxies out to $\sim$5000 km/s.}

\end{deluxetable}
}

In this paper,
we have derived the value of the Hubble constant using the I--band Tully--Fisher
relation.
The BVRIH$_{-0.5}$ TF relations were calibrated using ground--based
photometry data from Macri et al. (2000) and Aaronson, Huchra and Mould (1982,1986),
and Cepheid distances published as part of the series in Papers I through XXIII of this series.
These were then applied to the published cluster surveys
of Giovanelli et al. (I--band), Bothun et al. (B--band and V--band),
and Aaronson et al. (H--band) to derive the value of H$_0$.
The clusters extend out to $V_{cmb} \simeq 10,000$ km s$^{-1}$.
Taking the weighted average of Hubble constants derived using four different
surveys, we obtained $H_0 = 71 \pm 4$ (random) $ \pm 7$ (systematic) km s$^{-1}$ Mpc$^{-1}$.

Table 7 lists published H$_0$ values derived using the TF relation.
Focusing on the values published during past five years, the estimates
vary from the low 50s to the low 80s.
There are mainly three reasons for this variation.
First is the calibration of the TF relation.  Mould et al. (1996) showed
that doubling the number of calibrators changed the zero point
of the TF relation by nearly 0.3 mag, consequently lowering the value of
H$_0$ by 15\%.  Another example arises when comparing our I--band H$_0$ value
with that of Giovanelli (1997).
As shown in Table 5, using Giovanelli et al's extinction correction method,
and applying that to all the clusters (no velocity cutoff), we obtained
H$_0 = 76 \pm 3$ km s$^{-1}$ Mpc$^{-1}$, whereas Giovanelli (1997)
reported H$_0 = 70 \pm 5$ km s$^{-1}$ Mpc$^{-1}$.  Even though these two
values agree with each other within 1$\sigma$, the difference
is attributed to a different calibration.
Giovanelli (1997) used a smaller sample of 11 galaxies, instead of 21 that we used
in this paper.   Using the calibrators of Giovanelli (1997),
we do in fact retrieve the same H$_0$ value.
Increasing the number of calibrators again by 70\% (from 11 to 19) 
contributes considerably to the change in the value of H$_0$.
Furthermore, the same argument applies for the difference between
the values of H$_0$ obtained by Aaronson et al (1980) and this paper
which used the same set of clusters.  The calibration adopted by
the former consisted only of two galaxies, M31 and M33.
The decrease in H$_0$ is almost entirely due to the fact
that these two galaxies lie well below the
TF relation, thereby systematically yielding a lower H$_0$ value.

The second reason behind the various inconsistent estimates is the treatment
of the cluster population incompleteness bias.  Sandage, Tammann, Federspiel and
collaborators apply a large incompleteness bias to their B--band TF application,
largely based on the assumption that the TF intrinsic dispersion in B--band
is as large as 0.7 mag.
We have shown that the intrinsic TF dispersion is only $\sim$0.30 mag,
even in B--band.  In fact, there is no anti--correlation between the wavelength
and the dispersion, as some papers had suggested earlier (e.g. Pierce \& Tully 1988).

Third, some of the B--band TF distances to clusters have been measured
with respect to the Virgo cluster whose distance modulus was assumed to be
$31.60 \pm 0.08$ mag (Tammann 1999).  When there are nearly two dozen TF calibrators,
there is no need to estimate TF distances based on one cluster distance.

Results from the KLUN (Kinematics of the Local Universe)
survey have given H$_0$ values around 50--55 km s$^{-1}$ Mpc$^{-1}$
(cf. Theureau et al. 1997).  This is a B--band, 21cm survey of $\sim$2000
spiral galaxies out to 2000---3000 km s$^{-1}$.  
The latest results from this survey (Ekholm et al. 1999)
reported H$_0$ = 52 $\pm$ 5 km $^{-1}$ Mpc$^{-1}$ 
using the inverse diameter B--band TF relation, and 53 $\pm$ 6 using the inverse
magnitude B--band TF relation.
The KLUN database includes field galaxies, which need to be corrected for
Malmquist bias. They also found that their calibrator sample and the field
sample followed different diameter slopes, and subsequently corrected for this effect.
Both of these corrections decrease the value of H$_0$.
Our TF analyses suggested that the cluster incompleteness bias was minimal,
requiring a correction of only 2--3\% in distance.
Furthermore, we did not find a significant difference between the slopes
of the calibrator and cluster samples.

We have also explored how systematic uncertainties can
affect estimates of H$_0$.  The main results are summarized as follows:

(1) Intrinsic dispersions of the TF relations are $\sigma \simeq 0.19-0.25$ mag for $BVRIH_{-0.5}$.
The dispersion does not seem to decrease significantly with wavelength as reported earlier.

(2) No color dependence of the TF relation was observed.

(3) The cluster population incompleteness bias was found to be small,  typically
$\sim 0.05$ mag at most per cluster in distance modulus.

(4) The dependence of H$_0$ on the TF slope is negligible.

(5) We found that our estimate of H$_0$ is not sensitive to the choice of
internal extinction correction.  The estimates of H$_0$ using three different
extinction correction methods all agree well within 1$\sigma$ uncertainties.

(6) A very small dependence of the value of H$_0$\ on the metallicity dependence
of the Cepheid PL relation was observed.  When a metallicity dependence of $\gamma = 0.24
\pm 0.16$ mag dex$^{-1}$ is adopted, H$_0$ decreases by 3\%.

(7) H$_0$ is not sensitive to the velocity field model adopted.
We have used the CMB reference frame as a default case in measuring H$_0$.
However, we have also examined the case in which the flow field was estimated
using a three--attractor model. That yielded an H$_0$ 
that is exactly the same as that obtained using the CMB velocities.

(8) The larger two databases, based on the I--band and H--band surveys, which are likely
more statistically reliable than the other two, give Hubble constants that are
different by 10\%.  
We suggest that this difference arises from systematic photometric
differences between the calibrators and cluster galaxies in H$_{-0.5}$ magnitudes,
which are circular aperture magnitudes.
This clearly needs to be revisited in the future, possibly by obtaining
deep H--band total magnitudes consistently for both the calibrators and cluster
galaxies.

\medskip
\medskip
\medskip
It is a pleasure to thank Martha Haynes for generously 
providing us with the 21cm line profiles.  
Also we would like to thank Riccardo Giovanelli, Brent Tully,
Mike Pierce and Greg Bothun for discussions.
We also thank Jeff Mader for assembling
the initial database and references for our calibrator line-widths.
S.S. acknowledges support from NASA through the 
Long Term Space Astrophysics Program, NAS-7-1260.
L.M. acknowledges partial support by AURA through Gemini Fellowship No. GC--1003--95.
L.F. acknowledges support by NASA through Hubble Fellowship 
grant HF-01081.01-96A awarded by the Space Telescope Science Institute, which is operated
by the Association of Universities for Research in Astronomy, Inc., for NASA under contract NAS 5-26555.
S.M.G.H. and P.B.S. acknowledge support from a NATO collaborative
research grant (CRG960178).
The work presented in this paper is based on observations with the NASA/ESA
Hubble Space Telescope, obtained by the Space Telescope Science Institute,
which is operated by AURA, Inc. under NASA contract No. 5--26555.  Support for
this work was provided by NASA through grant GO--2227--87A from STScI.
This research has made use of the NASA/IPAC Extragalactic Database (NED) 
which is operated by the Jet Propulsion Laboratory, Caltech, under contract 
with the National Aeronautics and Space Administration.

\newpage
\appendix
\section{A.  Internal Extinction Corrections}

We briefly review three different methods for correcting
for  internal extinction.  The ``internal extinction'' correction
here strictly refers to the relative correction of spiral galaxy
magnitudes to the face--on orientation; we do not apply an absolute, 
total extinction correction to any of our galaxies.

The first such correction to consider is that of Han (1992). 
Using a sample of 284 spiral galaxies in I--band, he presented
a set of parametric expressions as follows:
\begin{equation}\label{eqn:han_extinction}
A^{Han}_{int,I} = \left\{ \begin{array}{ll}
	-0.73 \log (b/a) & \mbox{T} \leq 3  \\
	-0.90 \log (b/a) & \mbox{T} = 4, 5  \\
	-0.51 \log (b/a) & \mbox{T} > 6  
	\end{array}
\right.
\end{equation}
where $(b/a)$ is the minor--to--major axis ratio, expressed by 
$(1-e)$.

G97 examined again several hundred Sbc-Sc
galaxies and derived a more complicated set of parametric
formulae for internal extinction corrections which depended
on both the inclination and the size of the galaxies.
The basic idea behind their correction is that the light traverses
a longer path in a larger galaxy, increasing the internal extinction;
the larger the galaxy is, the more correction is needed.
The internal extinction expression given by G97 is:
\begin{equation}
A^{G97}_{int,I} = \gamma (\log W) \log (a/b) 
\end{equation}
where $\gamma$ is parametrized as a function of linewidth,
whose values are given in Table 3 of G97 and vary between 0.57 and 1.70.
In addition to the size--dependent correction, G97 suggests a constant
offset for galaxies of earlier types.  
\begin{equation}
\beta = \left\{ \begin{array}{ll}
	0.10  & \mbox{T} = 1, 2 \\
	0.32  & \mbox{T} = 3 \\
	0.0   & \mbox{otherwise}.
	\end{array}
\right.
\end{equation}
The total extinction is thus given by $A^{G97}_{int} + \beta$.

%The third type of extinction correction is given by Tully et al. (1998:
%hereafter T98)
%who investigated the multi--color observations of a total of
%87 spiral galaxies in Pisces and Ursa--Major clusters.
%Their study supported the G97 conclusion that the internal extinction
%is dependent on the size of the galaxy, as inferred by its linewidth:

%  APPENDIX A

\section{B. Inclinations}

In Table~B1, we summarize the inclination information
compiled from previously published data for each galaxy.
The columns include: (1) name of the galaxy; (2) photometric
inclination angles published with references indicated by
numbers in brackets; and (3) kinematical inclination
angles.

{\scriptsize
\begin{deluxetable}{lll}
\tablewidth{0pc}
\tablecaption{\bf Inclination Angles}
\tablehead{
\colhead{NGC} &
\colhead{Photometric\tablenotemark{a}} & 
\colhead{Kinematical\tablenotemark{a}}\\
\colhead{} &
\colhead{(Degrees)} &
\colhead{(Degrees)} 
}
\startdata
NGC~224 & 72 ({\it 1,5}), 78 ({\it 2,3,4}) & 75--78 ({\it 18}) \nl
NGC~598 & 51 ({\it 1}), 55 ({\it 3}), 54 ({\it 2}) & 55 ({\it 6}) \nl
NGC~925 & 55 ({\it 7}), 52 ({\it 8,9}), 51 ({\it 10}), 54 ({\it 11}), 57 ({\it 2}) & 55 ({\it 12}) \nl
NGC~1365 & 57 ({\it 1,13}), 61 ({\it 14}), 63 ({\it 11}), 44 ({\it 15}) & 40 ({\it 16}) \nl
NGC~1425 & 62 ({\it 1}), 66 ({\it 15}), 64 ({\it 17}) &  \nl
NGC~2090 & 59 ({\it 1}), 60 ({\it 13}), 62 ({\it 17}) & \nl
NGC~2403 & 53 ({\it 1}), 55 ({\it 3}), 60 ({\it 4}), 61 ({\it 2}) & 60 ({\it 12}), 50 ({\it 6}) \nl
NGC~2541 & 58 ({\it 1}) & \nl
NGC~3031 & 59 ({\it 1,9}) & 58 ({\it 6}) \nl
NGC~3198 & 65 ({\it 1,5,3}) & 70 ({\it 12}) \nl
NGC~3319 & 56 ({\it 1,19}) & 59 ({\it 26})\nl
NGC~3351 & 47 ({\it 1,20}) & 34 ({\it 6}), 40 ({\it 21}) \nl
NGC~3368 & 45 ({\it 1}) & \nl
NGC~3621 & 50 ({\it 1,13}), 62 ({\it 3}), 65 ({\it 11}) & 65 ({\it 22}) \nl
NGC~3627 & 62 ({\it 1}) &  65 ({\it 6}) \nl
NGC~4414 & 54 ({\it 1}) & 50--70 ({\it 23}) \nl
NGC~4535 & 43 ({\it 1}), 45 ({\it 5}) & 40 ({\it 24}) \nl
NGC~4536 & 64 ({\it 5}), 65 ({\it 7}), 69 ({\it 20}) &  \nl
NGC~4548 & 36 ({\it 1}) & 42 ({\it 24}) \nl
NGC~4725 & 45 ({\it 1}) & 53 ({\it 12}) \nl
NGC~7331 & 70 ({\it 1,5}), 71 ({\it 7,23}) & 75 ({\it 6}) \nl
\enddata
\tablenotetext{a}{
(1) Bottinelli et al. 1984,
(2) Shanks et al. 1997,
(3) Huchtmeier \& Richter 1988,
(4) Rood \& Williams 1993,
(5) Biviano et al. 1990,
(6) Garcia--Gomez \& Athanoussla 1991,
(7) Krumm \& Salpeter 1979,
(8) Davis et al. 1983,
(9) Giraud 1986,
(10) Giovanelli \& Haynes 1985,
(11) Tully 1988,
(12) Wevers et al. 1984,
(13) Reif et al. 1982,
(14) Schoniger \& Sofue, 1994,
(15) Bureau et al. 1996,
(16) Jorsater \& van Moorsel 1995,
(17) Mathewson et al. 1996,
(18) Corradi \& Capaccioli 1991, 
(19) Davis \& Seaquist 1983,
(20) Jackson et al. 1989,
(21) Buta 1988,
(22) Walsh 1999,
(23) Braine \& Combes 1993,
(24) Guhathakurta et al. 1988,
(25) Dumke et al. 1995.
(26) Moore \& Gottesman, 1998
}
\end{deluxetable}
}

\section{C. Quadratic Tully--Fisher Relations}

Several papers have investigated the possibility of a Tully--Fisher
relation being a quadratic relation (cf: Mould, Han and Bothun 1989).
Pierce and Tully (1988) noted that the TF relation at faint end tends
to curve downward.  Giovanelli et al. (1997) however noted that their sample
of several hundred galaxies showed no such trend.
We have fitted a quadratic equation to each $BVRIH$ Tully--Fisher correlation
and obtained the following results where $w = \log W^c_{20\%} - 2.5$:

   21   -19.8380   -7.94259    8.11011
   17   -20.4028   -8.41765    6.44754
   21   -20.6891   -9.34836    9.29448
   21   -21.1989  -10.19832    11.5500
   20   -21.8610   -12.1775    12.7621

\begin{equation}
B^C_T =  -19.84  - 7.94 w + 8.11 w^2
\end{equation}
\begin{equation}
V^C_T =  -20.40 - 8.42 w + 6.45 w^2
\end{equation}
\begin{equation}
R^C_T =  -20.69 - 9.35 w + 9.29 w^2
\end{equation}
\begin{equation}
I^C_T = -21.20 - 10.20 w + 11.55 w^2
\end{equation}
\begin{equation}
H^C_{-0.5} = -21.86 - 12.18 w + 12.76 w^2
\end{equation}
In Figure 13, these fits are shown by solid lines.  In comparison,
linear fits (Equations 6--10) are drawn by dashed lines.

If one were to use the quadratic equation instead, distances to individual 
galaxies may change slightly, especially for those with
large linewidth ($\log W > 2.6$) or smaller linewidths ($\log W < 2.3$).
However, its effect on the value of H$_0$ should be negligible for
two reasons: (1) the distance to the cluster should not change in the mean,
and (2) we are restricting the cluster sample galaxies
to those in the range of $2.3 \leq \log W \leq 2.6$.
Because the number statistics are not sufficient to conclude whether the TF
relation is quadratic or not, and also because the calibrator sample spans
a rather small range of linewidths, we will not use above equations (C1) through
(C5) in our further analysis.

\newpage

\include{fig}

\end{document}

%% file: fig.tex
\begin{figure}
\figurenum{1}
\plotone{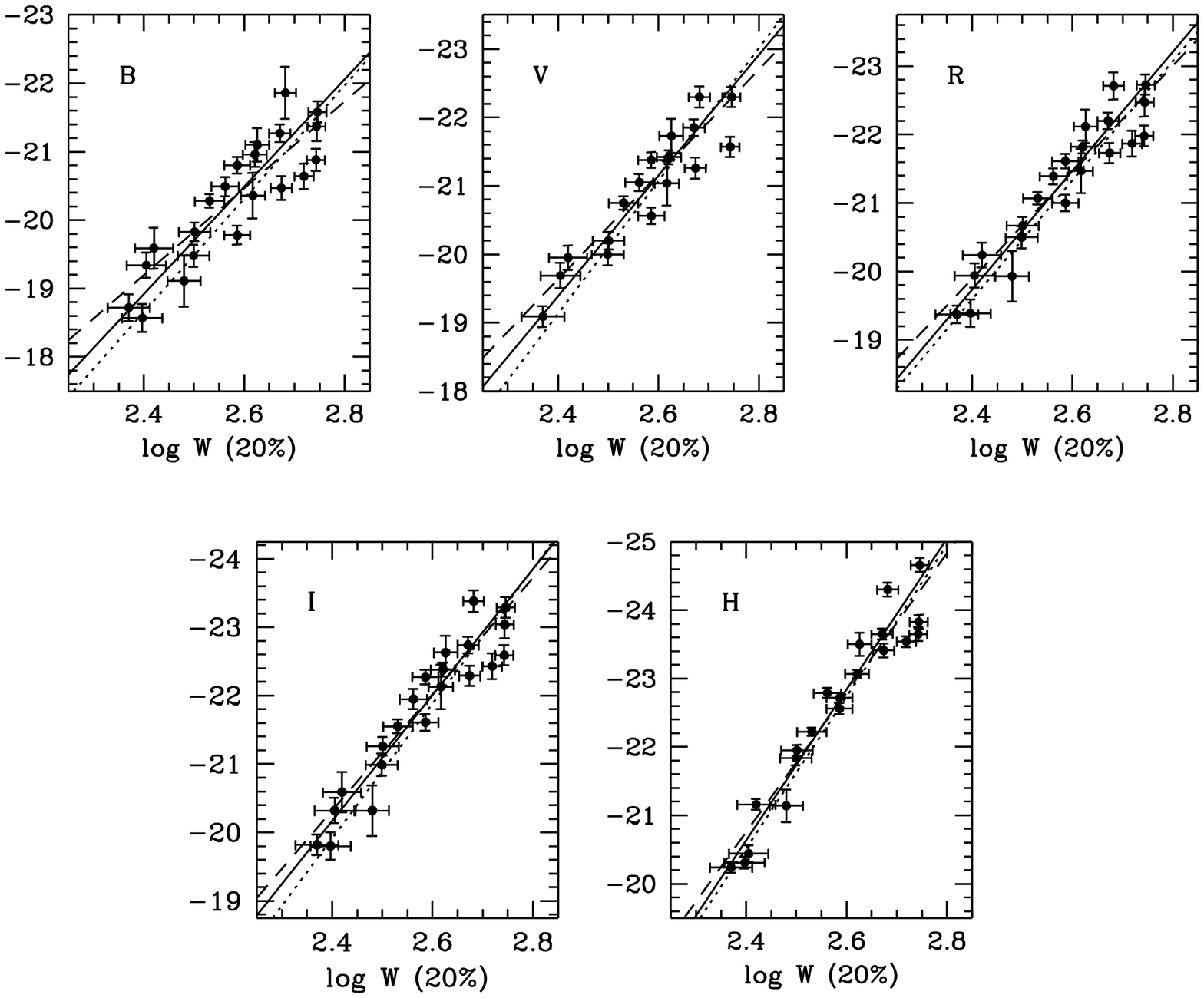}
\figcaption{BVRIH$_{-0.5}$ Tully--Fisher Relations for spiral 
galaxies with Cepheid distances, using 20\% linewidth.
Solid lines represent the bivariate fits, while
the dotted and dashed lines represent inverse and direct fits.}
\end{figure}

\begin{figure}
\figurenum{2}
\plotone{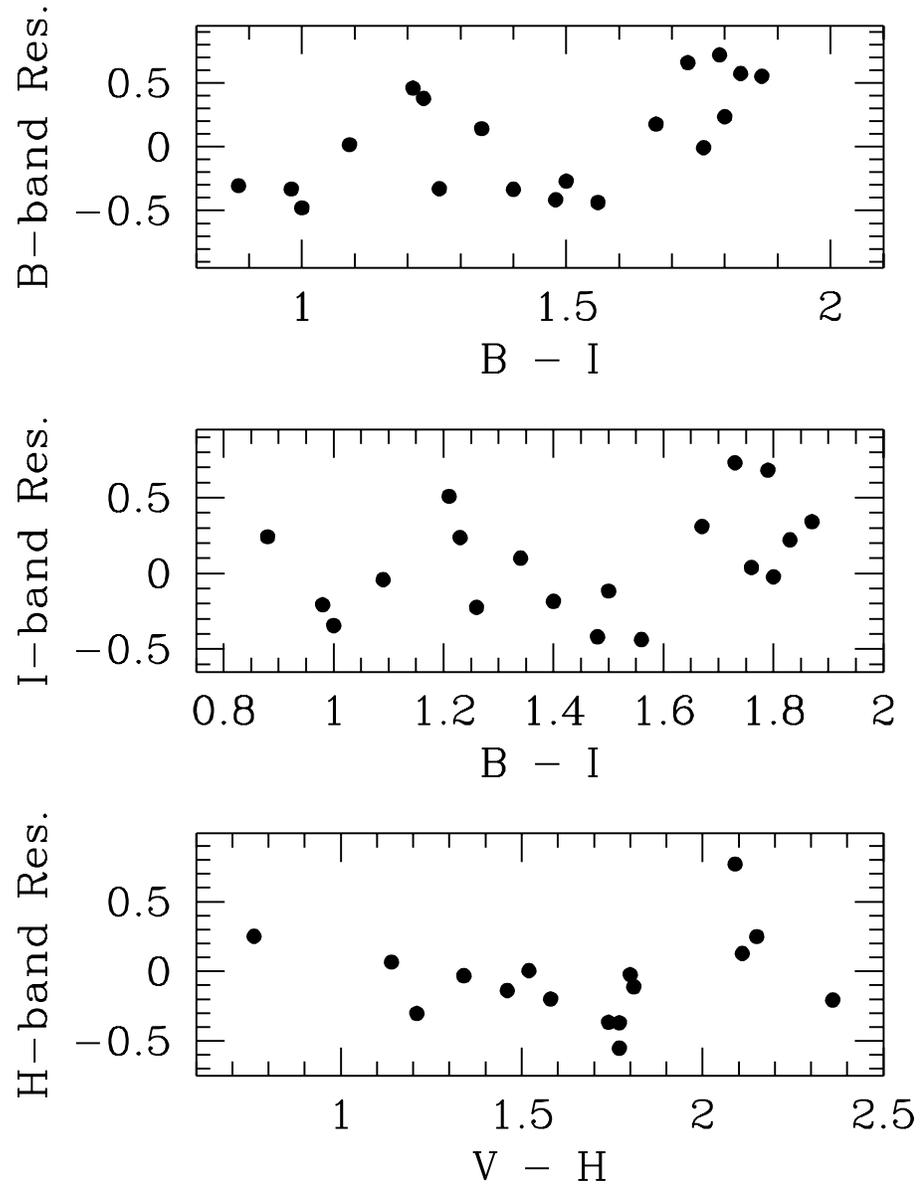}
\figcaption{Observed minus predicted absolute magnitude as a function
of color.  Although the sample is small, there is no obvious dependence
of the TF relation on color.}
\end{figure}

\begin{figure}
\figurenum{3}
\plotone{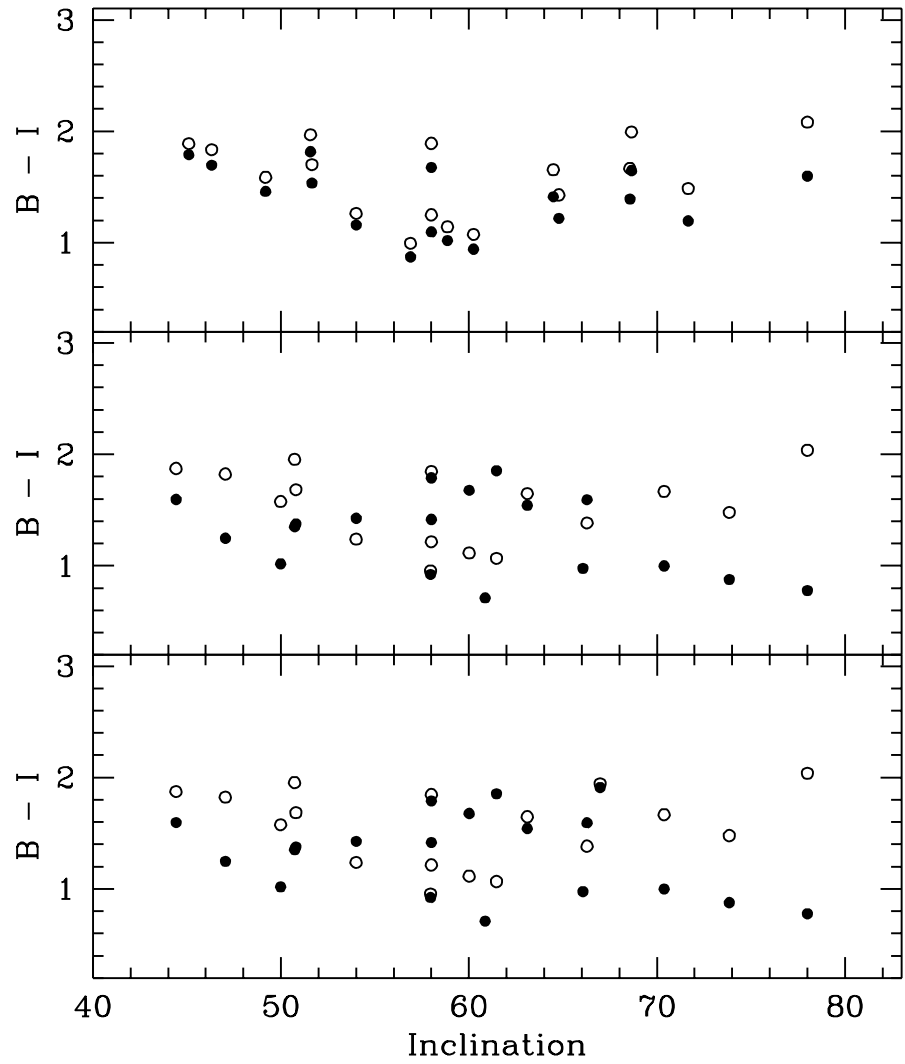}
\figcaption{Color dependence on inclination for TF calibrator galaxies.
Solid circles represent the correlations 
after correcting for the internal extinction using Tully et al.'s method
(top), Han's method (middle) and Giovanelli et al.'s method (bottom). 
Open circles are for the colors uncorrected for internal extinction.
If the internal  extinction correction were ``correct'', this correlation should
have zero slope.}
\end{figure}

\begin{figure}
\figurenum{4}
\plotone{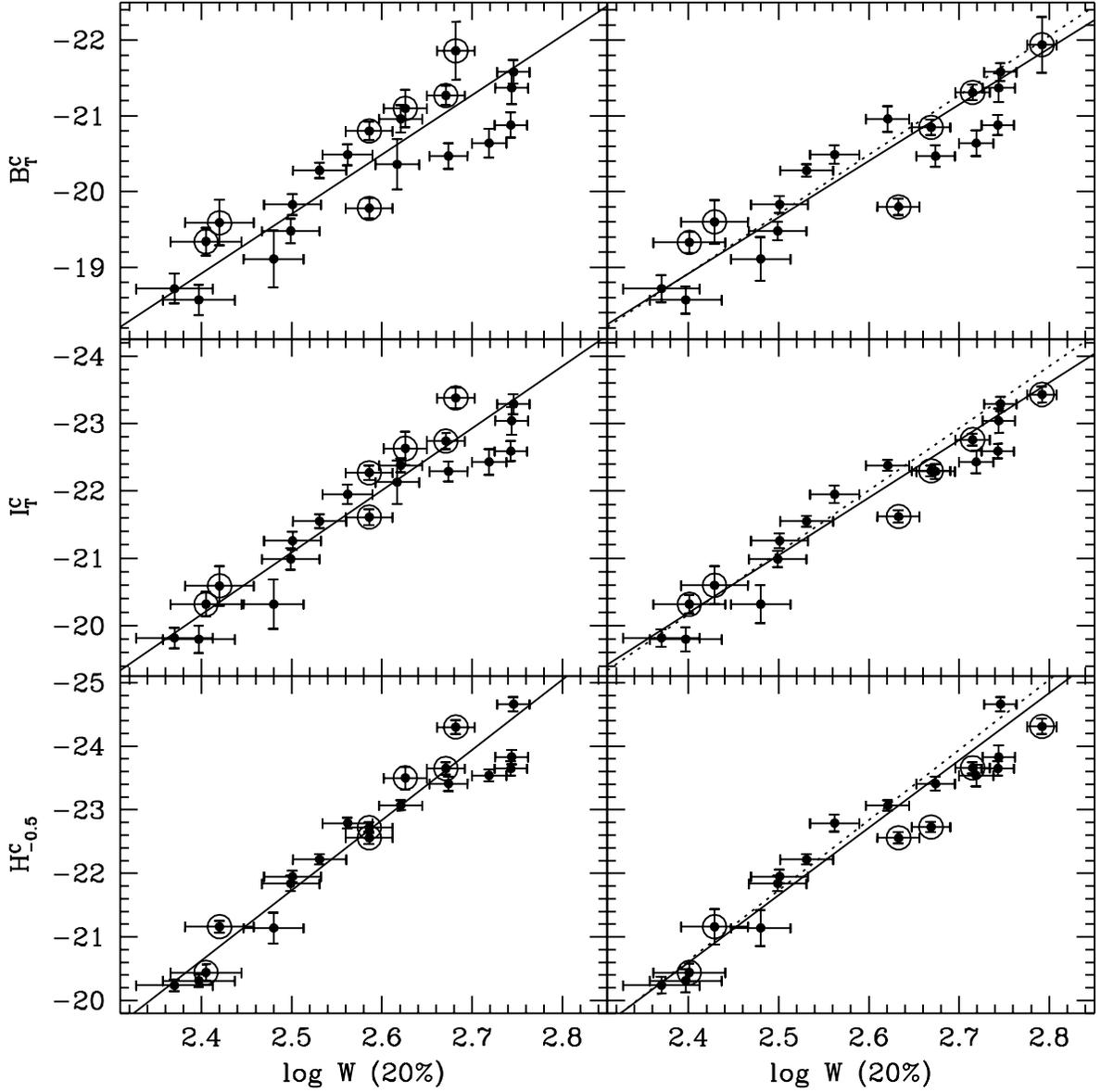}
\figcaption{{\it (left):}  B, I and H--band Tully--Fisher relations in which
the photometric inclination angles were used.
The barred galaxies are indicated by open circles.
The solid lines represent the TF fit through all the galaxies,
while the dashed lines the fit using only the barred galaxies.
{\it (right):}  B, I and H--band TF relations in which the
kinematical inclination angles were adopted for barred galaxies only.
Their TF fits are shown by solid lines.
The dotted lines are the fits to all the galaxies using the photometric
inclination angles (solid lines on the left plots).
The zero point change in, for example, the I--band TF relation is 0.01 mag.}
\end{figure}

\begin{figure}
\figurenum{5}
\plotone{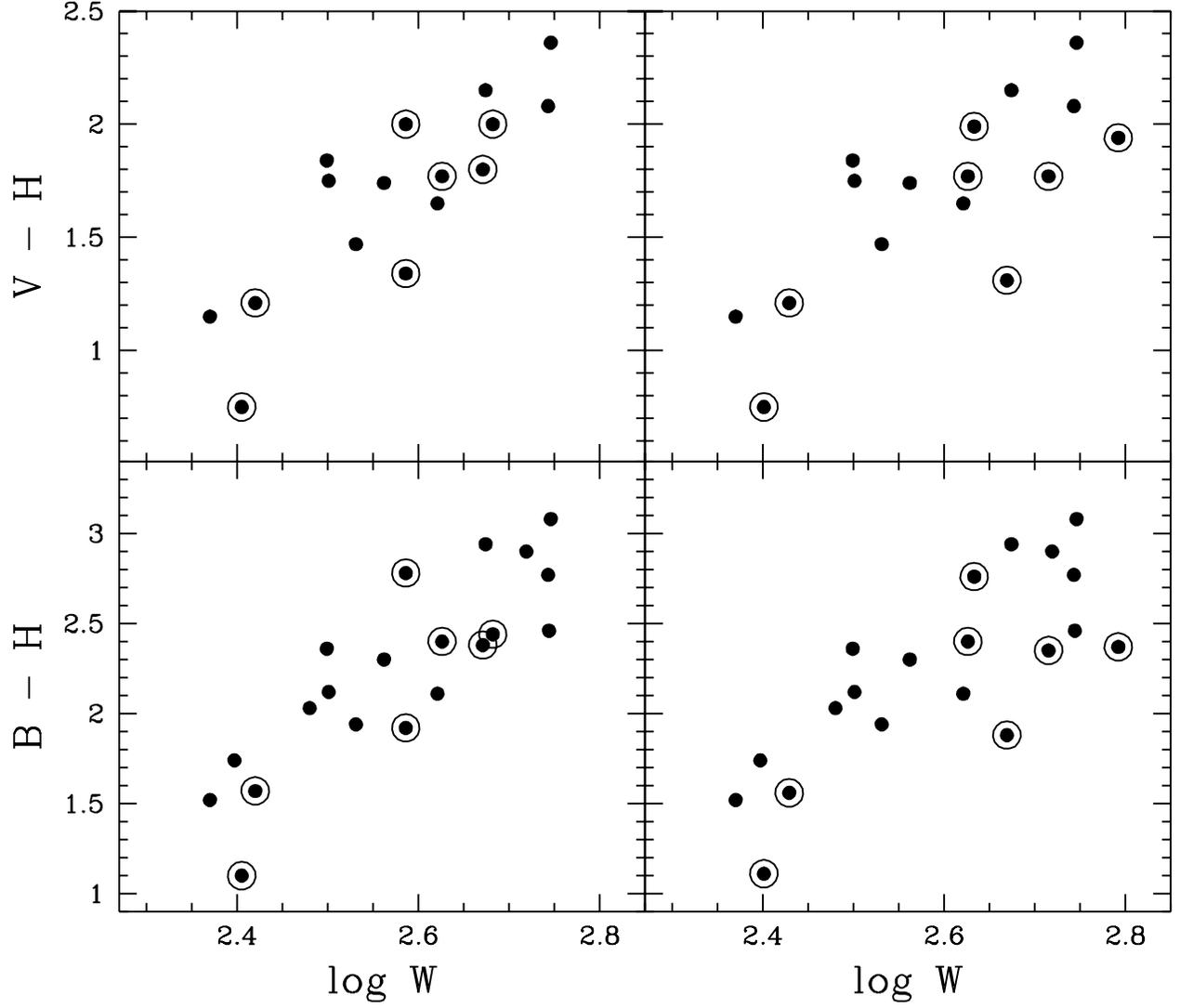}
\figcaption{The V--H and B--H color distributions of the local calibrators.
The barred galaxies are indicated by open circles.  On the left, the magnitudes
were corrected using the photometric inclination angles, while for those shown on
the right side, the kinematical inclination angles were substituted for the
barred galaxies.}
\end{figure}

\begin{figure}
\figurenum{6}
\plotone{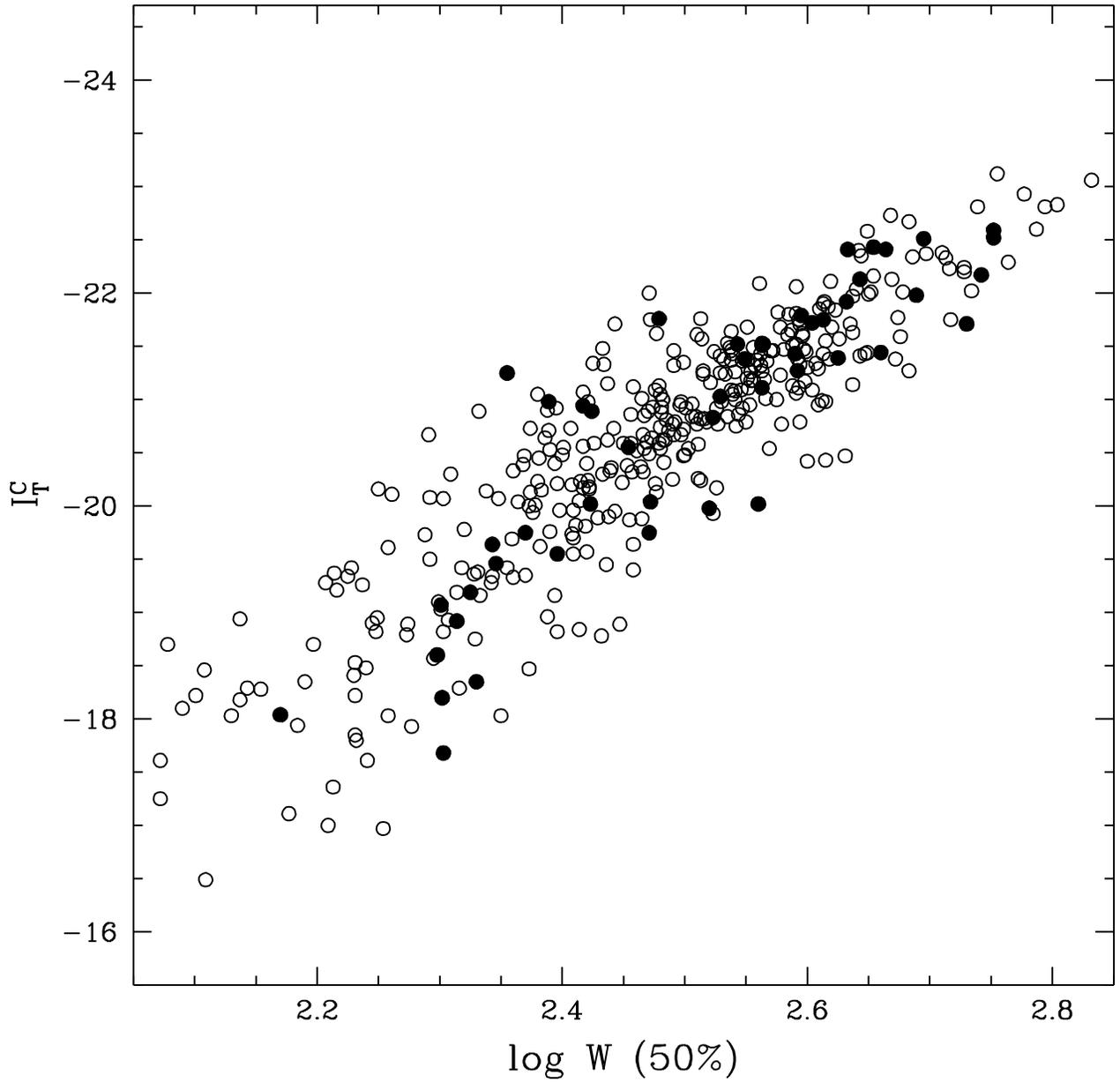}
\figcaption{I--band Tully--Fisher relation for the cluster survey galaxies.
The solid circles represent those galaxies classified as barred.}
\end{figure}

\begin{figure}
\figurenum{7}
\plotone{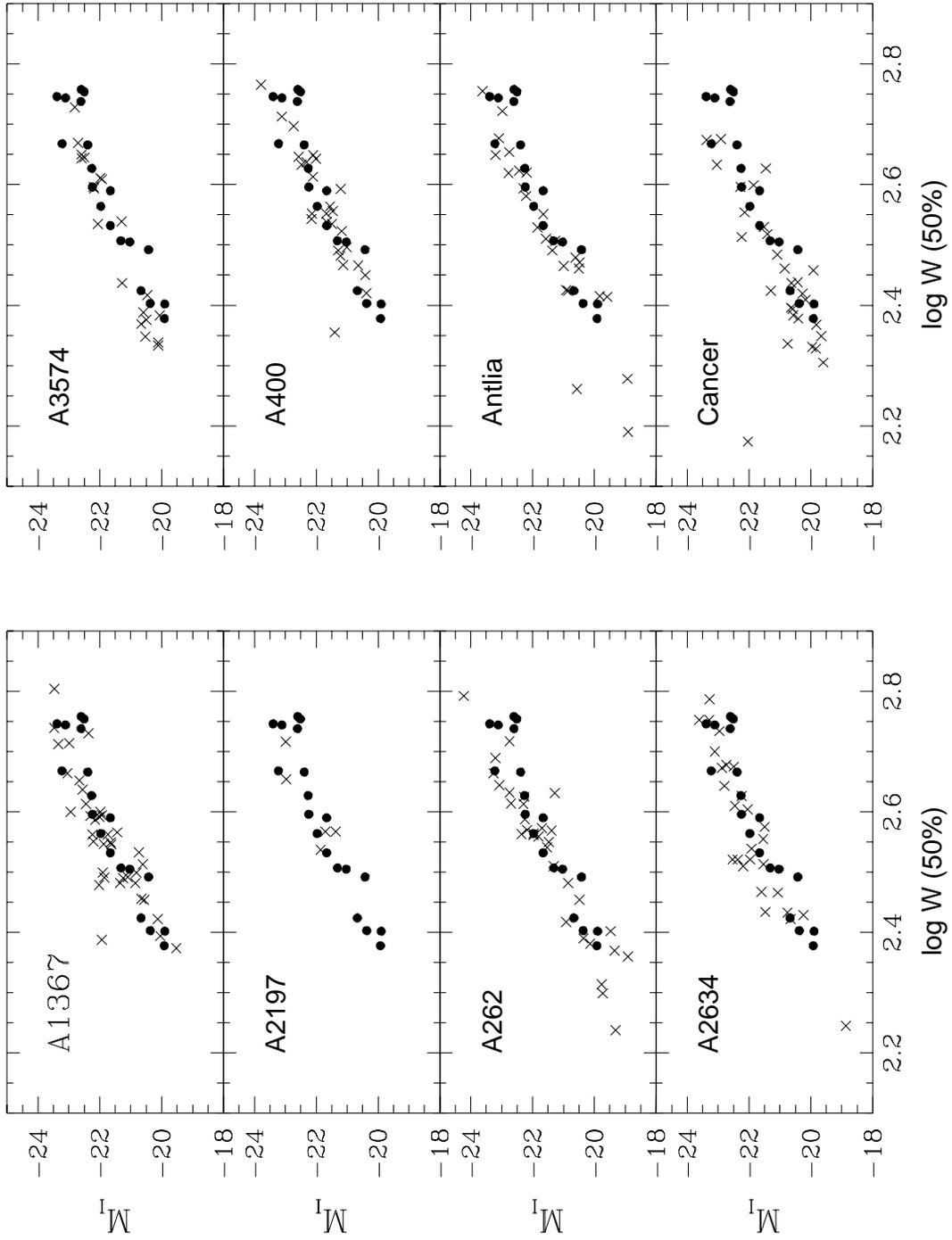}
\figcaption{ Examples of comparisons of the calibration galaxies (solid circles) 
with the cluster galaxies in the I--band survey (crosses).}
\end{figure}

\begin{figure}
\figurenum{8}
\plotone{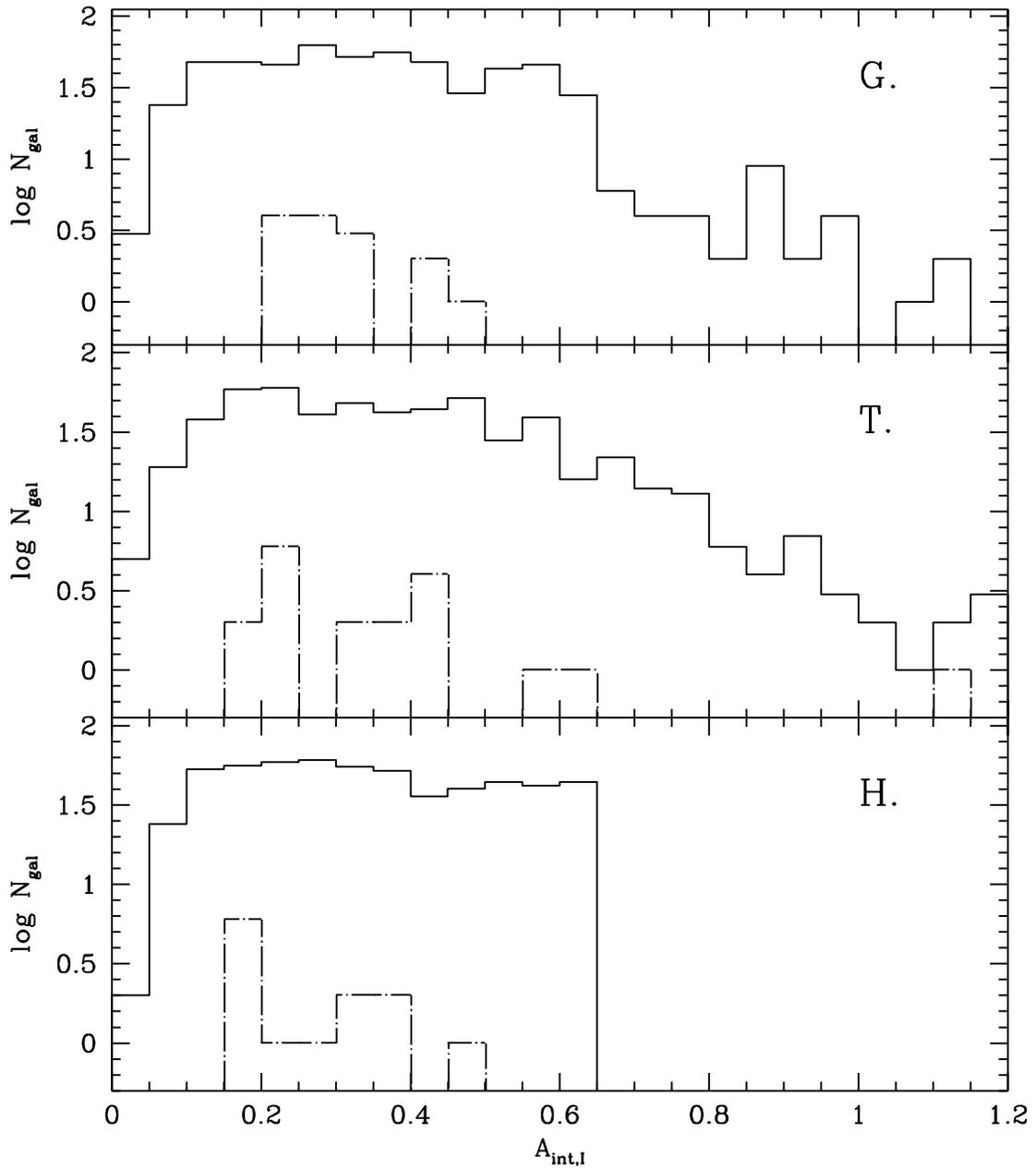}
\figcaption{Comparison of distributions of internal extinction corrections:
using Giovanelli et al (top), Tully et al. (middle) and Han's method (bottom).
The solid histograms represent the distributions for cluster galaxies, while
the dotted--dashed histograms are for the calibrators.  In our analysis, we
exclude any galaxies whose internal extinction exceeds 0.6 mag.}
\end{figure}

\begin{figure}
\figurenum{9}
\plotone{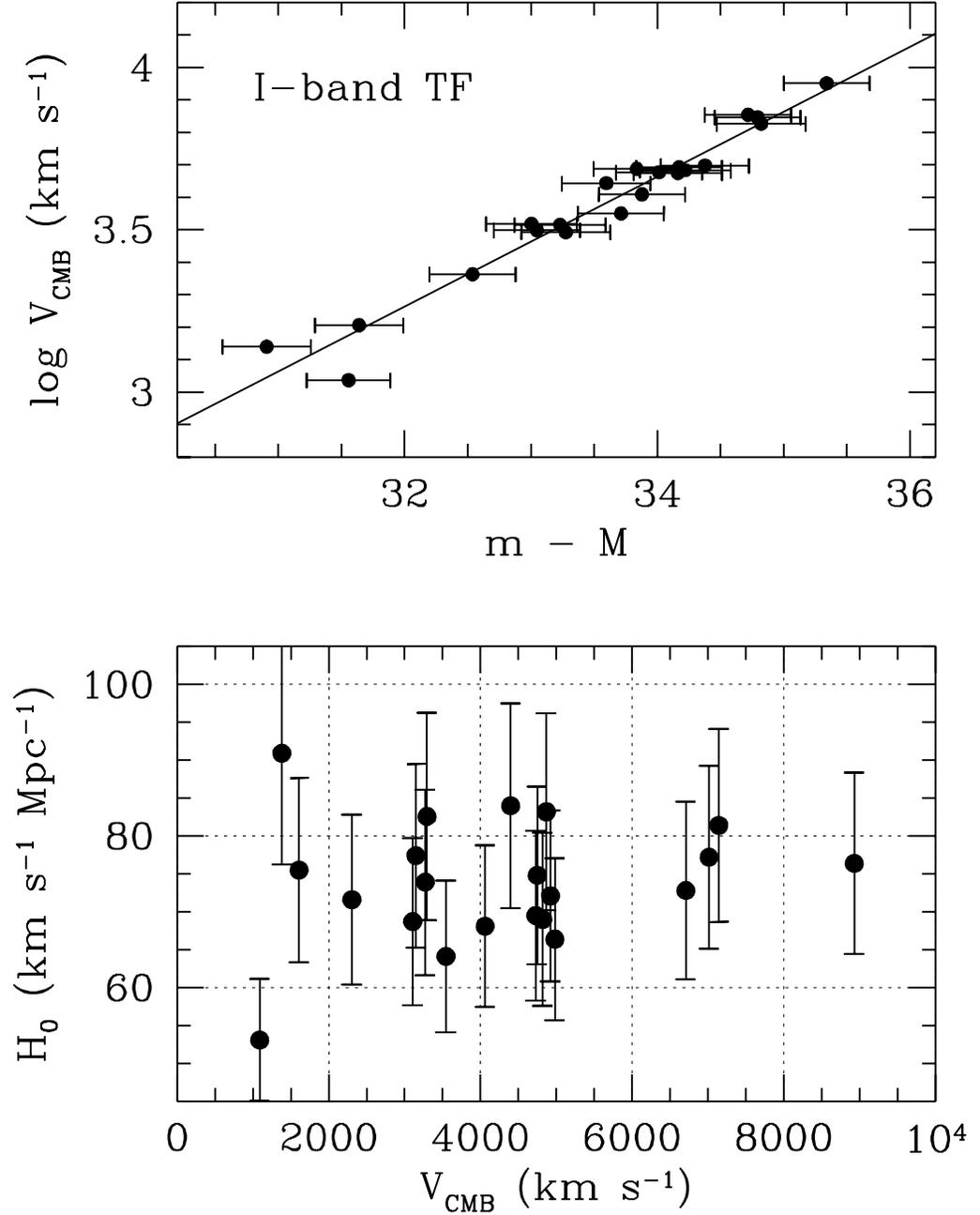}
\figcaption{A $\log D - \log V$ relation for I--band clusters
(top).  A straight line represents H$_0 = 73$ km s$^{-1}$ Mpc$^{-1}$.
The bottom figure shows H$_0$ as a function of velocity in the CMB
reference frame.  Taking the mean of clusters with $V_{cmb} \geq 2000$
km s$^{-1}$, we obtain H$_0$ = 73 $\pm$ 2 (random) km s$^{-1}$ Mpc$^{-1}$.}
\label{fig:iband_hubble_constant}
\end{figure}

\begin{figure}
\figurenum{10}
\plotone{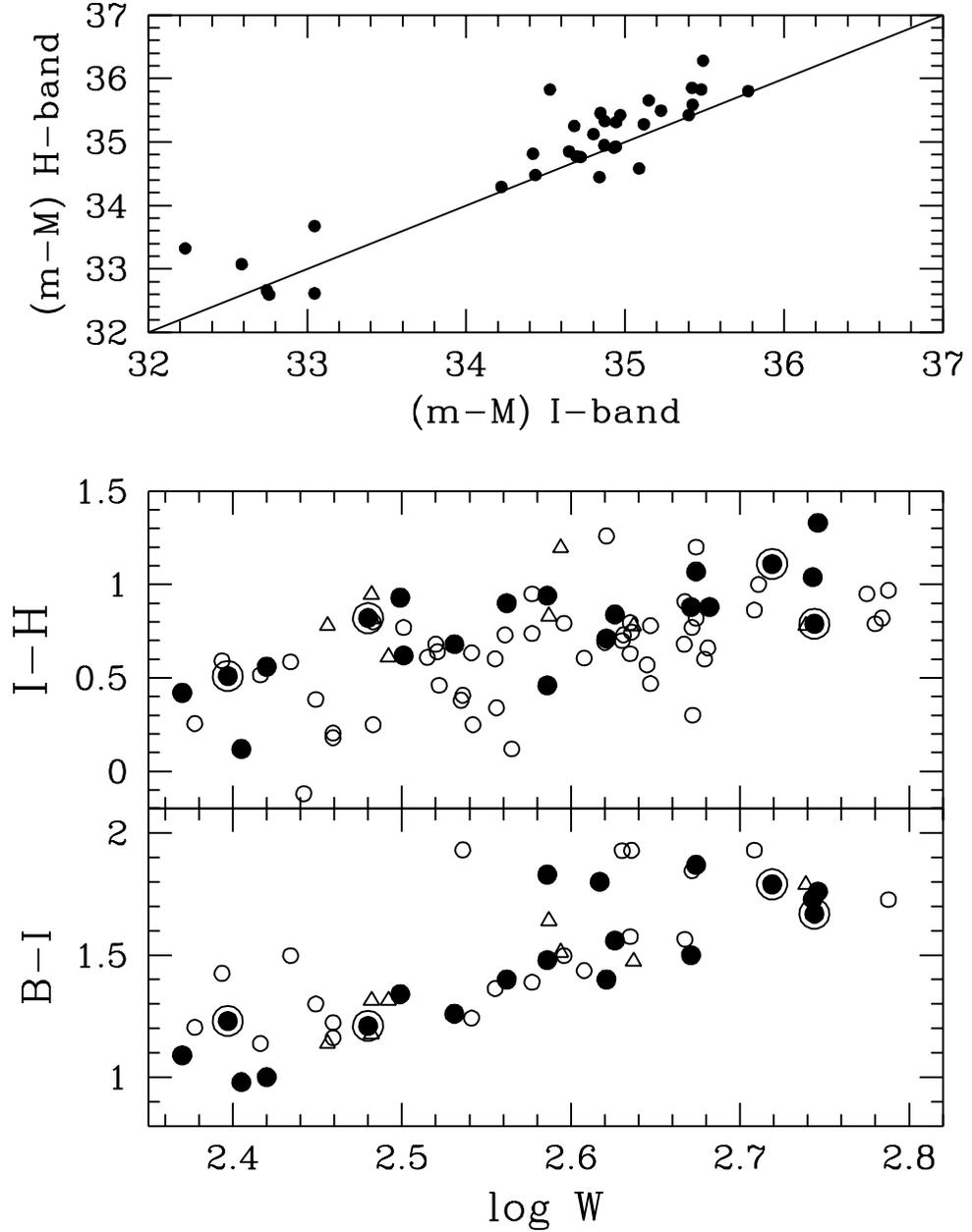}
\figcaption{{\it Top}: Comparison of distance moduli derived from the I--band
and H--band TF relations for galaxies that are included in both surveys.
The solid line represents an one--to--one correspondence and is not a fit
through the data.
{\it Middle}: The $I-H_{-0.5}$ color distribution as a function of logarithmic
linewidth for Virgo and Ursa Major galaxies (open squares) from Pierce \& Tully
(1988), and A1367, A400 and Coma galaxies whose photometric data (open triangles)
were compiled from the G97 and B85 databases.
The local calibrators are overplotted with solid circles. 
Note that the local calibrators are located on the red edge of the correlation
for the cluster galaxies.
{\it Bottom}: $B-I$ color as a function of linewidth for the same set of galaxies
as in the middle figure.
The local calibrators are  overplotted (solid circles). 
Four galaxies used by PT88  whose Cepheid 
distances were  measured from ground are indicated by solid circles 
overplotted with open circles. }
\end{figure}

\begin{figure}
\figurenum{11}
\plotone{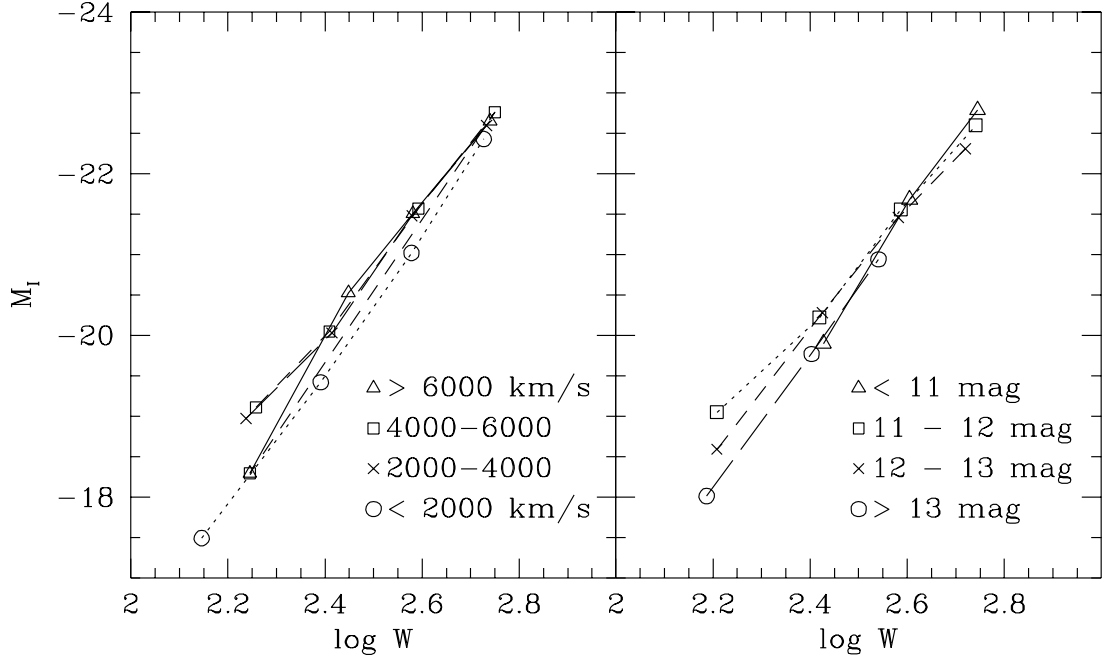}
\figcaption{I--band Tully--Fisher relations for samples grouped by velocities
in the CMB reference frame (left), and grouped by apparent magnitudes (right).
The absolute magnitudes were calculated assuming $H_0 = 100$km 
sec$^{-1}$ Mpc$^{-1}$ and that velocities give distances to first order.  
With an exception of the nearest sample, there seems to be no dependence
of Tully--Fisher zero point on the sample.  For the nearest sample ($<2000$kms$^{-1}$),
the offset is likely due to the fact that its velocity field is not characterized
well by the CMB model.}
\end{figure}

\begin{figure}
\figurenum{12}
\plotone{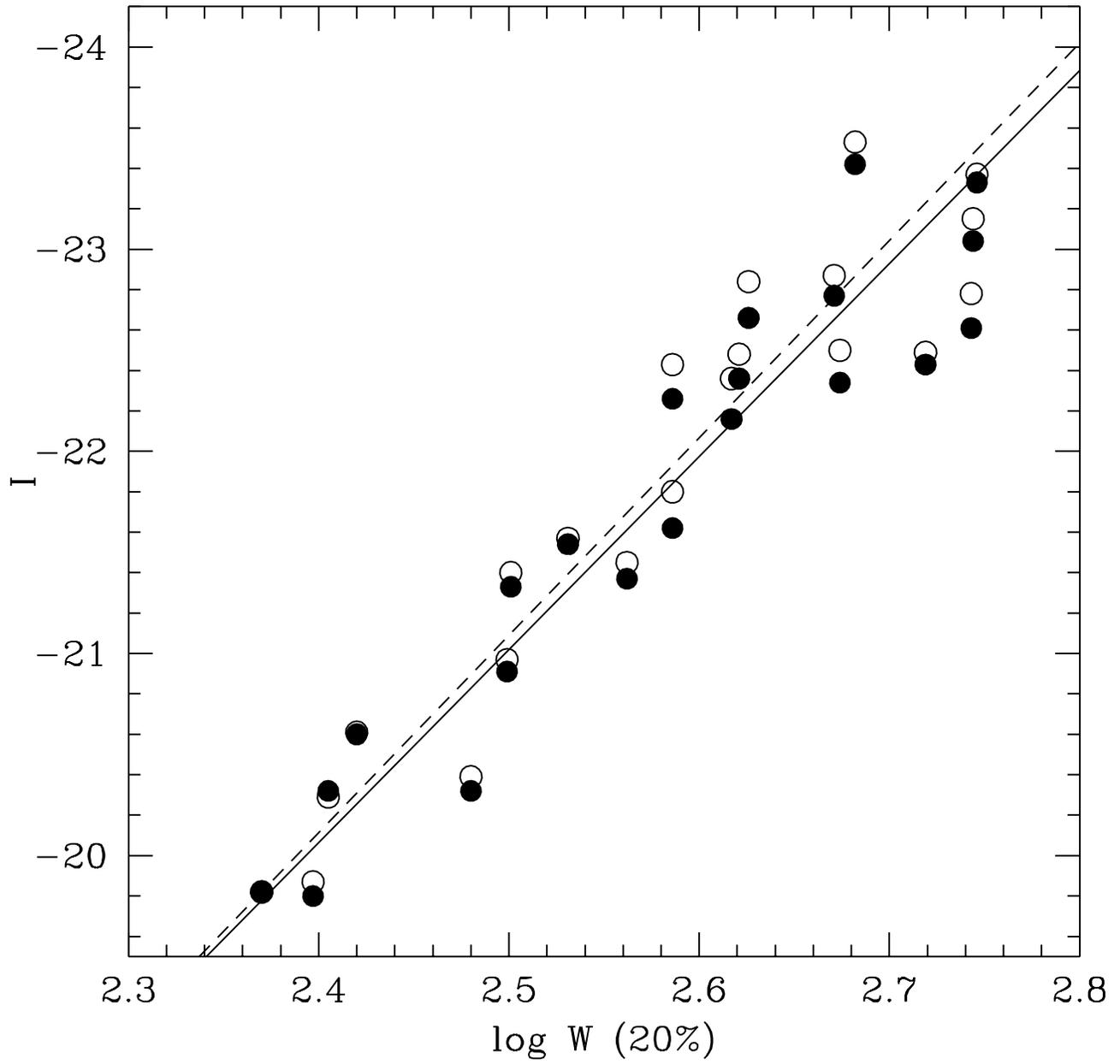}
\figcaption{I--band--$\log W_{20}$ Tully--Fisher relation (solid
circles).
Open circles present the luminosity--linewidth relation for same
galaxies, except that their distances have been adjusted to incorporate
the metallicity dependence of the Cepheid PL relation, using $\gamma = 
0.24$ mag dex$^{-1}$ (see text).}
\label{fig:tf.z.dependence}
\end{figure}

\begin{figure}
\figurenum{13}
\plotone{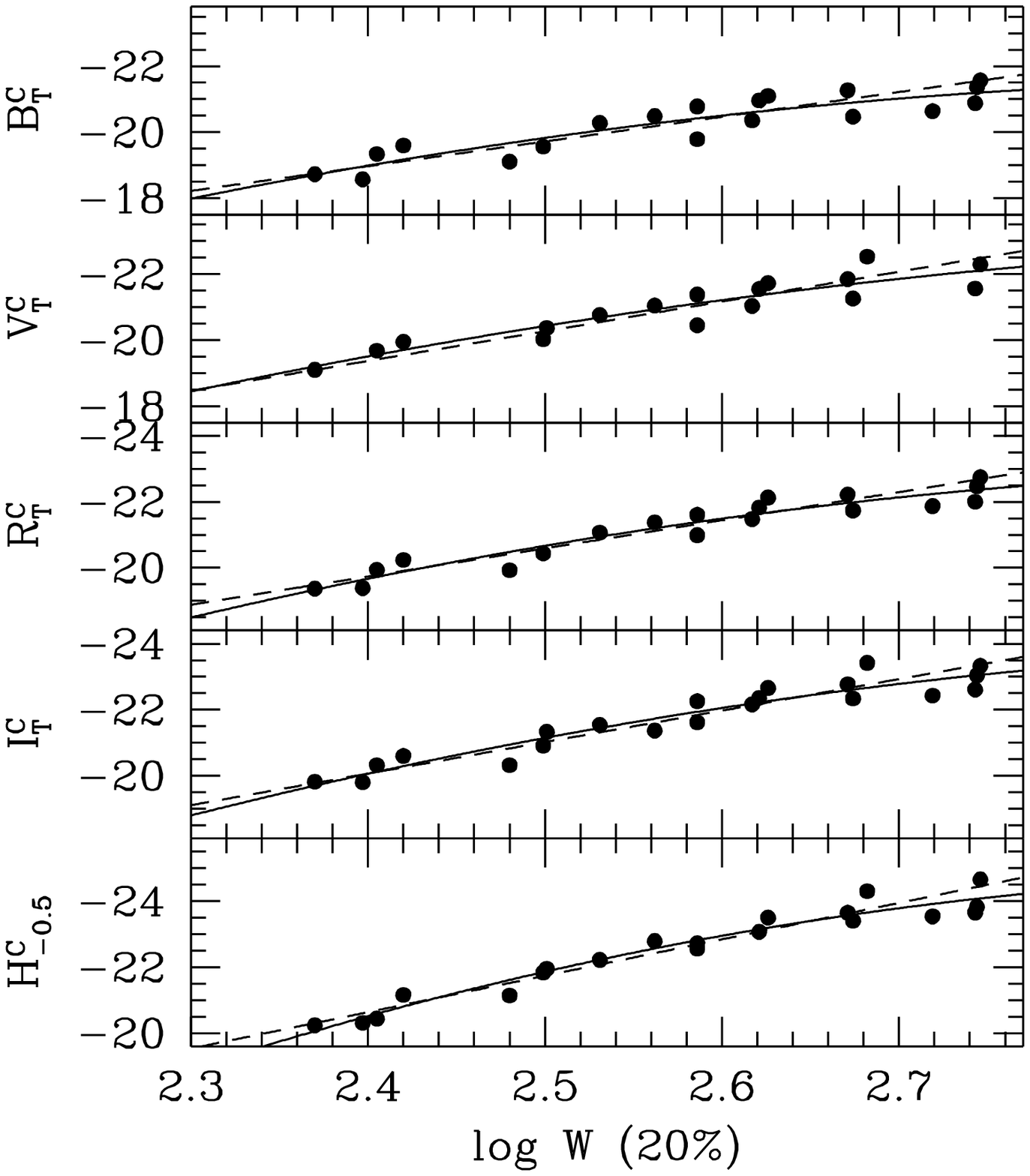}
\figcaption{The solid line in each Tully--Fisher relation represents a quadratic
fit while the dashed line is the linear relation (Equations 6 -- 10).
Applying the I--band quadratic TF relation to the cluster data, we obtain
H$_0$ = 75 $\pm$ 5 km s$^{-1}$ Mpc$^{-1}$.}
\end{figure}

%% file: preprint.bbl
\begin{thebibliography}{}

%\bibitem[]{} Aaronson, M., 1986, in ``Galaxy Distances and Deviations'', eds. B.F.Madore and
%R.B.Tully

\bibitem[]{} Aaronson, M., Mould, J.R., 1983, \apj, 265, 1

\bibitem[]{} Aaronson, M., Mould, J.R., Huchra, J.P., 1979, \apj, 229, 1

\bibitem[]{} Aaronson, M., Mould, J., Huchra, J., Sullivan, W.T., Schommer, R.A., 
Bothun, G.D., 1980, \apj, 239, 12

\bibitem[]{} Aaronson, M., Huchra, J.P., Mould, J.R., Tully, R.B.,
	Fisher, J.R., Vanwoerden, H., Goss, W.M., 
	Chamaraux, P., Mebold, U., Siegman, B., Berriman, G. and
	Persson, S.E., 1982, \apjs, 50, 241

\bibitem[]{} Aaronson, M., Bothun, G., Mould, J.R., Huchra, J., 
Schommer, R.A., \& Cornell, M.E., 1986, \apj, 302, 536

\bibitem[]{} Biviano, A., Giuricin, G., Mardirossian, F., Mezzetti, M.,
1990, \apjs, 74, 325

\bibitem[]{} Bernstein, G.M., Guhathakurta, P., Raychaudhury, S., Giovanelli, R., Haynes, M.P.,
Herter, T. \& Vogt, N.P., 1994, \aj, 107, 1962

%\bibitem[]{} Bothun, G.D., Aaronson, M., Schommer, B., Huchra, J., Mould, J., 1984, \apj, 278, 475

\bibitem[]{} Bothun, G.D., Aaronson, M., Schommer, B., Mould, J., Huchra, J., Sullivan, W.T., 1985, \apjs, 57, 423

\bibitem[]{} Bottinelli, L., Goughenheim, L., 1976, \aap, 51, 275

\bibitem[]{} Bottinelli, L., Goughenheim, L., Paturel, G., de Vaucouleurs, G.,
1984, \aaps, 56, 381

\bibitem[]{} Bottinelli, L., Gouguenheim, L., Paturel, G. \& Teerikorpi, P., 1988, \apj, 328, 4

\bibitem[]{} Braine, J. \& Combes, F., 1992, \aap, 264, 433

\bibitem[]{} Bureau, M., Mould, J.R., \& Staveley--Smith, L., 1996, \apj, 463, 60

%\bibitem[]{} Burstein, D., \& Heiles, C., 1982, \aj, 87, 1165

\bibitem[]{} Burstein, D., Willick, J.A., \& Courteau, S., 1995, in ``The
	Opacity in Spiral Disks'', ed. J.I.Davies and D. Burstein, Kluwer
	Academic Publishers

\bibitem[]{} Buta, R., 1988, \apjs, 66, 233

\bibitem[]{} Cardelli, J.A., Clayton, G.C. \& Mathis, J.S., 1989, \apj, 345, 245

\bibitem[]{} Corradi, R.L.M. \& Capaccioli, M., 1991, \aaps, 90, 121

%\bibitem[]{} Courteau, S., 1992, PhD Thesis, University of California, Santa Cruz

\bibitem[]{} Davis, L.E., \& Seaquist, E.R., 1983, \apjs, 53, 269

\bibitem[]{} de Jong, R.S., 1997, PhD Thesis, Kapteyn Astronomical Institute

\bibitem[]{} de Vaucouleurs, G., de Vaucouleurs, A., \& Corwin, J.R., 1976, Second
Reference Catalog of Bright Galaxies (RC2)

\bibitem[]{} Dekel, A., 1994, \araa, 32, 371

\bibitem[]{} Dumke, M., Krause, M., Wielebinski, R., \& Klein, U., 1995, \aap, 
	302, 691

\bibitem[]{} Ekholm, T., Teerikorpi, P., Theureau, G., Hanski, M., Paturel, G., Bottinelli, L.,
 \& Gouguenheim, L., 1999, \aap, in press

\bibitem[]{} Eisenstein, D.J. \& Loeb, A., 1996, \apj, 459, 432

%\bibitem[]{} Feast, M.W., 1994, \mnras, 266, 255

\bibitem[]{} Federspiel, M., Sandage, A. \& Tammann, G.A., 1994, \apj, 430, 29 (FST94)

\bibitem[]{} Federspiel, M., Tammann, G.A., \& Sandage, A., 1998, \apj, 495, 115
%\bibitem[]{} Ferrarese, L. et al., 1996, \apj, 464, 568

\bibitem[]{} Ferrarese, L. et al., 1998, \apj, 507, 655

\bibitem[]{} Ferrarese, L. et al., 2000a, submitted %[popII]

\bibitem[]{} Ferrarese, L. et al., 2000b, submitted %[data]

\bibitem[]{} Fouque, P., Bottinelli, L., Gouguenheim, L., Paturel, G., 1990, \apj, 349, 1

\bibitem[]{} Freedman, W.L., 1990, \apj, 355, L35

\bibitem[]{} Freedman, W.L. et al. 1994, \apj, 427, 628  %M81

\bibitem[]{} Freedman, W.L. et al. 2000, \apj, in preparation %[finale paper]

%bibitem[]{} Freudling, W., Da Costa, L.N., Wegner, G., Giovanelli, R., Haynes, M.P.,
%alzer, J.J., 1995, \aj, 110, 920

\bibitem[]{} Fukugita, M., Okamura, S., Tarusawa, K., Rood, H.J., Williams, B.A., 1991, \apj, 376, 8

\bibitem[]{} Garcia--Gomez \& Athanoussla, 1991, \aaps, 89, 159

%\bibitem[]{} Gavazzi, G., 1993, \apj, 419, 469

\bibitem[]{} Gibson, B.G. et al. 1999, \apj, in press %[n4725]

\bibitem[]{} Gibson, B.G. et al. 2000, submitted %[SNIa]

\bibitem[]{} Giovanelli, R., 1997, in ``The Extragalactic Distange Scale'',
	M.Livio, M.Donahue, and N.Panagia, eds., Cambridge University Press

\bibitem[]{} Giovanelli, R., \& Haynes, M.P., 1985, \aj, 90, 2445
 
\bibitem[]{} Giovanelli, R., Haynes, M.P., Salzer, J.J., Wegner, G., Da Costa, L.N.,
Freudling, W., 1994, \aj, 107, 2036   %extinction I

\bibitem[]{} Giovanelli, R., Haynes, M.P., Herter, T., Vogt, N.P., Wegner, G.,
Salzer, J.J., Da Costa, L.N., Freudling, W., 1997a, \aj, 113, 22  % data 

\bibitem[]{} Giovanelli, R., Haynes, M.P., Herter, T., Vogt, N.P., Da Costa, L.N.,
Freudling, W., Salzer, J.J., Wegner, G., 1997b, \aj, 113, 53    % CPIB paper

\bibitem[]{} Giovanelli, R., Haynes, M.P., Salzer, J.J., Wegner, G., DaCosta, L.N.,
Freudling, W., 1998, \aj, 116, 2632

\bibitem[]{} Giraud, E., 1986, \aap, 164, 17

\bibitem[]{} Gould, A., 1993, \apj, 412, L55

\bibitem[]{} Gould, A., 1994, \apj, 426, 542

\bibitem[]{} Graham, J.A. et al., 1997, \apj, 477, 535

\bibitem[]{} Graham, J.A. et al., 1999, \apj, 516, 626

\bibitem[]{} Guhathakurta, P., van Gorkom, J.H., Kotanyi, C.G., \& Balkowski, C., 1988, \aj, 96, 851

\bibitem[]{} Han, M., 1991, PhD Thesis, California Institute of Technology

\bibitem[]{} Han, M., 1992, \apj, 391, 617

%\bibitem[]{} Han, M. \& Mould, J.R., 1990, \apj, 360, 448

\bibitem[]{} Han, M. \& Mould, J.R., 1992, \apj, 396, 453

%\bibitem[]{} Han, M. et al., 1998, preprint

\bibitem[]{} Holmberg, E. 1958, Medd Lund Obs II, No. 136

%\bibitem[]{} Huchra, J.P. et al. 1999, in preparation  %flow model paper

\bibitem[]{} Huchtmeier, W.K. \& Richter, O.G., 1988, \aap, 203, 237

\bibitem[]{} Hughes, S.M.G. et al., 1998, \apj, 501, 32

\bibitem[]{} Jackson, J.M., Snell, R.L., Ho, P.T.P., \& Barrett, A.H., 1989, \apj, 337, 680

\bibitem[]{} Jorsater, S. \& van Moorsel, G., 1995, \aj, 110, 2037

%\bibitem[]{} Kelson, D.D. et al., 1996, \apj, 463, 26

\bibitem[]{} Kelson, D.D. et al., 1999, \apj, 514, 614

\bibitem[]{} Kelson, D.D. et al., 2000, \apj, submitted %[fundamental plane]

%\bibitem[]{} Kennicutt, R.C., Freedman, W.L. \& Mould, J.R., 1995, \aj, 110, 1476

\bibitem[]{} Kennicutt, R.C., 1999, \araa, 36

\bibitem[]{} Kennicutt, R.C. et al., 1998, \apj, 498, 181

\bibitem[]{} Kent, S.M., 1983, \apj, 266, 562

\bibitem[]{} Kraan--Korteweg, R.G., Cameron, L.M., and Tammann, G.A., 1988,
	\apj, 331, 620

\bibitem[]{} Krumm, N., \& Salpeter, E.E., 1979, \aj, 84, 1138

\bibitem[]{} Landy, S.D., \& Szalay, A.S. 1992, \apj, 391, 494

\bibitem[]{} Lauer, T.R. \& Postman, M., 1994, \apj, 425, 418

\bibitem[]{} Lynden--Bell, D., Burstein, D., Davies, R.L., Dressler, A., Terlevich, R.J.,
\& Wegner, G., 1988, \apj, 326, 19

\bibitem[]{} Macri, L.M. et al., 1999, \apj, 521, 155

\bibitem[]{} Macri, L.M. et al., 2000, \apj, in preparation  %[TF data paper]

\bibitem[]{} Madore, B.F. \& Freedman, W.L., 1991, \pasp, 103, 933

\bibitem[]{} Madore, B.F. et al., 1999, \apj, 515, 29 %[fornax implication]

%Freedman, W.L., Silbermann, N., Harding, P., Huchra, J., Mould, J.R.,
%	Graham, J.A., Ferrarese, L., Gibson, B.K., Han, M., Hoessel, J.G., Hughes, S.M.,
%	Illingworth, G.D., Kelson, D., Phelps, R., Sakai, S., 
%	Stetson, P., 1998, \apj, in press %(Fornax implication

\bibitem[]{} Mathewson, D.S. \& Ford, V.L., 1996, \apjs, 107, 97

\bibitem[]{} Moore, E.M. \& Gottesman, S.T., 1998, \mnras, 294, 353

\bibitem[]{} Mould, J.R., Aaronson, M., Huchra, J.P., 1980, \apj, 238, 458 

%\bibitem[]{} Mould, J.R., Kennicutt, R.C. \& Freedman, W.L., 1998

\bibitem[]{} Mould, J.R., Han, M \& Bothun, G., 1989, \apj, 347, 112

\bibitem[]{} Mould, J.R., Sakai, S., Hughes, S. \& Han, M., 1996, Space Telescope Institute
Symposium Series, Vol. 10, p. 158. ed. Livio, Donahue, and Panagia

\bibitem[]{} Mould, J.R. et al., 1998, \apj, in press %(N1425)

\bibitem[]{} Mould, J.R. et al., 2000, \apj, submitted %CtC paper

\bibitem[]{} \"Opik, E., 1922, \apj, 55, 406

\bibitem[]{} Peletier, R.F., \& Willner, S.P., 1993, \apj, 418, 626

\bibitem[]{} Phelps, R.L. et al., 1998, \apj, 500, 763

%\bibitem[]{} Pierce, M.J., 1994, \apj, 430, 53

\bibitem[]{} Pierce, M.J., \& Tully, B.R., 1988, \apj, 330, 579

\bibitem[]{} Pierce, M.J. \&  Tully, B.R., 1992, \apj, 387, 47   %absolute calibration, M31,M33 mags

%\bibitem[]{} Prosser, C. et al., 1998, \apj, in press
 
\bibitem[]{} Rawson, D.M. et al. 1997, \apj, 490, 517

\bibitem[]{} Reif, K., Mebold, U., Goss, W.M., van Woerden, H., \& Siegman, B., 1982, \aaps, 50, 451

\bibitem[]{} Riess, A.G., Press, W.H., \& Kirshner, R.P., 1995, \apj, 445, 91

\bibitem[]{} Roberts, M.S., 1969, \aj, 74, 859

\bibitem[]{} Roberts, M.S., 1978, in Vol. IX, {\it Stars and Stellar Systems}, ed. by
A. Sandage, M.Sandage, \& J. Kristian, Chicago: U. of Chicago Press, P.309

\bibitem[]{} Rood, H.J., \& Williams, B.A., 1993, \mnras, 263, 211

%\bibitem[]{} Saha, A., Labhardt, L., Schwengeler, H., Macchetto, F.D., Panagia, N.,
%	Sandage, A. and Tammann, G.A., 1994, \apj, 425, 14

%\bibitem[]{} Saha, A., Sandage, A., Labhardt, L., Schwengeler, H., 
%	Tammann, G.A., Panagia, N. and Macchetto, F.D.,
%	1995, \apj, 438, 8

\bibitem[]{} Saha, A., Sandage, A., Labhardt, L., Tammann, G.A.,
	Macchetto, F.D. and Panagia, N., 1996, \apj, 466, 55

\bibitem[]{} Saha, A., Sandage, A., Tammann, G.A., Labhardt, L., Macchetto, F.D. \& Panagia, N., 1999, \apj, in press

%\bibitem[]{} Saha, A., Sandage, A., Labhardt, L., Tammann, G.A., 
%	Macchetto, F.D. and Panagia, N., 1996, \apjs, 107, 693

%\bibitem[]{} Saha, A., Sandage, A., Labhardt, L., Tammann, G.A., 
%	Macchetto, F.D. and Panagia, N., 1997, \apj, 486, 1

\bibitem[]{} Sakai, S. et al. 1999, \apj, in press

\bibitem[]{} Sandage, A., 1994, \apj, 430, 13

\bibitem[]{} Sandage, A., Tammann, G.A., 1976, \apj, 210, 7

\bibitem[]{} Sandage, A., Tammann, G.A., 1984, \nat, 307, 326

%\bibitem[]{} Sandage, A., Tammann, G.A. \& Federspiel, M., 1995, \apj, 452, 1

%\bibitem[]{} Sandage, A., Saha, A., Tammann, G.A., Panagia, N. and Macchetto, D.,
%	1992, \apjl, 401L, 7

%\bibitem[]{} Sandage, A., Saha, A., Tammann, G.A., Labhardt, L.,
%	Schwengeler, H., Panagia, N. and Macchetto, F.D.,
%	1994, \apjl, 423L, 13

\bibitem[]{} Schechter, P.L., 1980, \aj, 85, 801

\bibitem[]{} Schlegel, D.J., Finkbeiner, D.P., \& Davis, M. 1998, \apj, 500, 525

\bibitem[]{} Schoniger, F. \& Sofue, Y., 1994, \aap, 283, 21

\bibitem[]{} Silbermann, N.A. et al, 1996, \apj, 470, 1

\bibitem[]{} Silbermann, N.A. et al, 1999, \apj, 515, 1

\bibitem[]{} Silk, J., 1996, in ``the Universe at High z, Large Scale Structure
and the Cosmic Microwave Background'', E.Martinez--Gonzalez, \& J.-L. Sanz,
	eds., Springer--Verlag, in press

\bibitem[]{} Shanks, T., 1997, \mnras, 290, 77

\bibitem[]{} Strauss, M.A., \& Willick, J.A., 1995, Physics Reports, 261, 271

%\bibitem[]{} Tammann, G.A., 1996, \pasp, 108, 1083

\bibitem[]{} Tammann, G.A., 1999, IAU Symposium \#183, ``Cosmological Parameters and the
Expansion of the Universe''

\bibitem[]{} Tanvir, N.R., Shanks, T., Ferguson, C and Robinson, R.T., 1995,
Nature, 377, 27

\bibitem[]{} Teerikorpi, P., 1984, \aap, 141, 407

\bibitem[]{} Teerikorpi, P., 1987, \aap, 173, 39

%\bibitem[]{} Teerikorpi, P., 1990, \aap, 234, 1

\bibitem[]{} Teerikorpi, P., 1997, \araa, 35, 101

\bibitem[]{} Theureau, G., 1998, \aap, 331, 1

\bibitem[]{} Theureau, G., Hanski, M., Ekholm, T., Bottinelli, L., Gouguenheim, L.,
Paturel, G., Teerikorpi, P., 1997, \aap, 322, 730

\bibitem[]{} Tormen, G., Burstein, D., 1995, \apjs, 96, 123

%\bibitem[]{} Triay, R., Lachieze--Rey, M. \& Rauzy, S., 1994, \aap, 289, 19

%\bibitem[]{} Tully, B.R., 1998, in ``Post--Hipparcos Cosmic Candles'', eds. Caputo and Heck

\bibitem[]{} Tully, B.R., 1999, IAU Symposium \#183

\bibitem[]{} Tully, R.B. \& Fisher, J.R., 1977, \aap, 54, 661

\bibitem[]{} Tully, R.B. \& Fouque, P., 1985, \apjs, 58, 67

\bibitem[]{} Tully, R.B., Mould, J.R., Aaronson, M., 1982, \apj, 257, 527
 
\bibitem[]{} Tully, R.B., Pierce, M.J., Huang, J, Saunders, W., Verheijen, M.W.,
Witchalls, P.L., 1998, \aj, 115, 2264 (T98)

\bibitem[]{} Turner, A. et al., 1998, \apj, 505, 207

\bibitem[]{} Valentijn, E.A., 1990, \nat, 346, 153

\bibitem[]{} Visvanathan, N., 1982, Proc. ASA, 4, 419

\bibitem[]{} Visvanathan, N., 1983, \apj, 275, 430

\bibitem[]{} Walsh, W. 1999, private communications

\bibitem[]{} Wevers, B.M.H.R., Appleton, P.N., Davies, R.D., \& Hart, L., 1984, \aap, 140, 125

\bibitem[]{} Willick, J.A., 1991, PhD Thesis, University of California, Berkeley

\bibitem[]{} Willick J.A., Courteau, S., Faber, S.M., Burstein, D., 
	Dekel, A., \& Kolatt, T., 1996, \apj, 457, 460

\bibitem[]{} Yahil, A., Tammann, G.A. \& Sandage, A., 1977, \apj, 217, 903

\bibitem[]{} Yasuda, N., Fukugita, M., Okamura, S., 1997, \apjs, 108, 417


 
 
\end{thebibliography}
